\documentclass[twocolumn, twocolappendix]{aastex63}
	
\usepackage[encapsulated]{CJK}
\usepackage{ucs}
\usepackage[utf8x]{inputenc}
\usepackage{url}
\newcommand{\cntext}[1]{\begin{CJK}{UTF8}{bsmi}#1\end{CJK}}

\usepackage[italicdiff]{physics} 
\usepackage{amsmath}

\usepackage{graphicx}

\usepackage{braket}

\usepackage{xcolor}

\usepackage{upgreek}

\newcommand{\m}[1]{\underline{\underline{#1}}}

\newcommand{\coone}{\mbox{CO(1--0)}}

\newcommand{\ci}{\textsc{C\,i}}

\newcommand{\hi}{\textsc{H\,i}}
\newcommand{\hii}{\textsc{H\,ii}}

\shorttitle{Dust and chemistry co-evolution in dwarf galaxies}
\shortauthors{Hu et al.}

\begin{document}

\defcitealias{HSvD21}{HSvD21}


\title{Co-evolution of Dust and Chemistry in Galaxy Simulations with a Resolved Interstellar Medium}
\author[0000-0002-9235-3529]{Chia-Yu Hu (\cntext{胡家瑜})}
\affiliation{Max-Planck-Institut f\"{u}r Extraterrestrische Physik, Giessenbachstrasse 1, D-85748 Garching, Germany}
\affiliation{Department of Astronomy, University of Florida, 211 Bryant Space Science Center, Gainesville, FL 32611, USA}
\author{Amiel Sternberg}
\affiliation{School of Physics \& Astronomy, Tel Aviv University, Ramat Aviv 69978, Israel}
\affiliation{Center for Computational Astrophysics, Flatiron Institute, 162 5th Ave, New York, NY 10010, USA}
\affiliation{Max-Planck-Institut f\"{u}r Extraterrestrische Physik, Giessenbachstrasse 1, D-85748 Garching, Germany}
\author{Ewine F. van Dishoeck}
\affiliation{Max-Planck-Institut f\"{u}r Extraterrestrische Physik, Giessenbachstrasse 1, D-85748 Garching, Germany}
\affiliation{Leiden Observatory, Leiden University, P.O. Box 9513, NL-2300 RA Leiden, the Netherlands}
\correspondingauthor{Chia-Yu Hu}
\email{cyhu.astro@gmail.com}

\begin{abstract}

Nearby dwarf irregular galaxies
are ideal laboratories for studying the interstellar medium (ISM) at low metallicity,
which is expected to be common for galaxies at very high redshift 
being observed by the James Webb Space Telescope.
We present the first
high-resolution ($\sim 0.2$~pc)
hydrodynamical simulations of an isolated low-metallicity ($0.1~Z_\odot$) dwarf galaxy 
coupled with a time-dependent chemistry network and a dust evolution model
where dust is locally produced and destroyed by various processes.
To accurately model carbon monoxide (CO),
we post-process the simulations with a detailed chemistry network
including the time-dependent effect of molecular hydrogen (H$_2$).
Our model successfully reproduces the observed star formation rate and {\coone} luminosity ($L_{\rm CO}$).
We find that dust growth in dense gas is required to reproduce the observed $L_{\rm CO}$ 
as otherwise CO would be completely photodissociated.
In contrast, the H$_2$ abundance is extremely small and is insensitive to dust growth,
leading to a CO-to-H$_2$ conversion factor 
that is only slightly higher than
the Milky Way value
despite the low metallicity.
An observationally inferred dust-to-gas ratio is thus underestimated
if adopting the metallicity-dependent CO-to-H$_2$ conversion factor.
The newly-produced dust in dense gas mixes with the ISM through supernova feedback
without being completely destroyed by sputtering,
which leads to galactic outflows 20\% -- 50\% dustier than the ISM,
providing a possible source for intergalactic dust.

\end{abstract}

\section{Introduction} \label{sec:intro}

The formation and evolution of galaxies are critically controlled by
how stars form and how they affect the gas cycle in and around galaxies via stellar feedback \citep{Somerville2015, Naab2017}.
Over the last decade,
the cold, star-forming gas 
(dominated by molecular hydrogen, H$_2$) 
in galaxies from the local Universe 
to ``cosmic noon'' at redshift $z \sim 2$ 
has been systematically quantified by submillimeter and far-infrared (FIR) telescopes, 
leading to a physical picture 
where 
galaxies grow primarily
by gas accretion onto the rotationally-supported disks, 
which fuels star formation in their interstellar medium (ISM) 
regulated by feedback across cosmic time (see \citealp{Tacconi2020} and references therein).
The molecular gas mass inferred by dust-based methods is broadly consistent with 
the conventional method based on carbon monoxide (CO),
strengthening the robustness of the results.

However,
while significant progress has been made 
in understanding the evolution of star-forming gas in galaxies of solar or slightly sub-solar metallicity, 
little is known about the ISM
of low-metallicity galaxies ($Z \lesssim 0.1 Z_\odot$).
In the era of James Webb Space Telescope (JWST),
a deep understanding of the ISM chemistry at low metallicity 
is urgently needed as we begin to observe galaxies in the very early Universe.
Indeed,
\citet{CurtisLake2022} recently reported 
four galaxies at extremely high redshift ($z\sim 10 -13$) discovered by JWST,
all of which have metallicity of $Z < 0.1 Z_\odot$.

On the other hand,
nearby dwarf irregular galaxies 
provide a unique laboratory to study
the chemical properties and observational signatures of the star-forming gas at comparably low metallicity in great detail thanks to their proximity,
even though their galaxy properties (such as mass, size, surface density, etc.)
may be different from their high-redshift counterparts.
It has long been recognized that
CO emission 
in these galaxies
tends to be extremely faint and 
becomes undetected when the metallicity drops below $0.2 Z_\odot$ \citep{Leroy2005, Schruba2012, Madden2020}.
This has changed with the advent of the Atacama Large Millimeter/submillimeter Array (ALMA)
thanks to its extremely high sensitivity.
\citet{Rubio2015} observed the Wolf-Lundmark-Melotte (WLM) dwarf galaxy 
and detected the first CO emission at a metallicity of $0.1 Z_\odot$.
However,
the molecular gas mass still cannot be robustly determined
due to the highly uncertain CO-to-H$_2$ conversion factor. 
Similarly,
the dust-based method suffers from 
the uncertainty in
the assumed dust-to-gas ratio (DGR)
that has been shown to scale super-linearly with metallicity in this regime,
but the exact scaling relation is still uncertain (see, e.g., \citealp{RemyRuyer2014, DeVis2019}).
In fact,
the detection of CO in the WLM galaxy is rather surprising
given the extremely low DGR expected at this metallicity.

Recent
hydrodynamical simulations coupled with time-dependent chemistry
have achieved the required numerical resolution of a few parsecs (pc) to directly resolve feedback from individual supernova (SN) explosions in isolated dwarf galaxies
\citep{Hu2016, Hu2017, Hu2019, Emerick2019, Lahen2020, Hislop2022, Whitworth2022, Whitworth2022a, Steinwandel2022a, Steinwandel2022, Katz2022, Lahen2022}.
This is an important milestone as it avoids the need for
sub-grid prescriptions for SN feedback 
one of the major uncertainties in cosmological simulations.
Coincidentally,
H$_2$ can be resolved at a similar resolution \citep{Gong2018},
at least at solar metallicity.
However,
proper modeling of CO requires a significantly higher resolution of $\sim 0.1$~pc \citep{Seifried2017, Hu2021a}
as CO typically exists in dense, spatially compact gas.
Lagrangian codes are therefore particularly suitable for such a task
thanks to their built-in adaptive spatial resolution
that can more easily reach sub-pc scales.
Meanwhile,
the carbon network responsible for CO formation is much larger than the hydrogen network \citep{Sternberg1995},
leading to a significant computational overhead if coupled with simulations,
compromising the achievable resolution.
To address this dilemma,
\citet{Hu2021a} introduced a hybrid approach
where the hydrogen network is solved on-the-fly 
to capture the time-dependent (non-steady-state) effect of H$_2$
while an accurate chemistry network including carbon chemistry is solved in post-processing.


Dust plays an important role in ISM chemistry.
The surfaces of dust grains are the main sites where H$_2$ formation occurs,
which then initiates the formation of other important molecules such as CO.
Furthermore,
dust provides shielding against the Lyman-Werner radiation that can photodissociate both H$_2$ and CO.
Observations of the Small and Large Magellanic Clouds (SMC and LMC)
have demonstrated that the spatially-resolved DGR can vary by almost an order of magnitude 
from region to region in low-metallicity galaxies,
indicating significant dust evolution \citep{RomanDuval2017, RomanDuval2022}.
However,
simulations that resolve the ISM to date have all assumed 
a constant DGR in both space and time, which is an over-simplification.

Cosmological simulations and isolated galaxy simulations with comparable resolution 
have started to include dust evolution models at different levels of sophistication
\citep{McKinnon2017, McKinnon2018, Aoyama2017, Aoyama2018, Li2019, Parente2022, Lewis2022, Romano2022, Lower2022}.
However, 
these simulations do not resolve either SN feedback or the density structure of the ISM,
forcing them to adopt dust evolution models in a sub-grid fashion.
For example,
dust destruction by SNe is generally based on results from 1D calculations in plane-parallel shocks,
which does not apply for more complex situations 
such as an inhomogeneous ISM or clustered SNe.
While this simplification is perhaps justified as SN feedback in these simulations is unresolved anyway,
a more accurate model is clearly desirable in resolved simulations.
\citet{Hu2019a} introduced a numerical method to 
directly simulate thermal and nonthermal sputtering of dust
designed for simulations with pc- or sub-pc resolutions.
However, the model did not include any dust production processes,
nor was it coupled with chemistry.



In this work,
we extend the dust evolution model in \citet{Hu2019a} 
by including dust growth in cold gas and dust production in asymptotic giant branch (AGB) stars.
We couple this with a time-dependent chemistry network and the ISM model
developed in \citet{Hu2016, Hu2017, Hu2021a}
and perform hydrodynamical simulations of an isolated dwarf galaxy similar to the WLM galaxy
at a mass resolution of $1~{\rm M_\odot}$ (spatial resolution $\sim 0.2$~pc).
To our knowledge,
this is the first ISM-resolved simulation coupled with chemistry and dust evolution 
(but see \citealp{Romano2022} for a model for significantly coarser resolutions).
This paper is organized as follows.
In Section \ref{sec:method}, we describe our numerical model.
In Section \ref{sec:results},
we study 
how dust evolution affects the ISM chemistry,
how the DGR is distributed in the ISM and galactic outflows,
and 
how the projected DGR varies 
with the gas surface density at different telescope beam sizes.
In Section \ref{sec:discussion},
we discuss the implications of our results for the observationally inferred DGR and the intergalactic dust.
In Section \ref{sec:summary},
we summarize our work.


\section{Numerical Methods} \label{sec:method}

\subsection{Gravity and hydrodynamics}

We use the public version of {\sc Gizmo} \citep{Hopkins2015},
a multi-solver code for hydrodynamics
that is built on the massively parallel TreeSPH code {\sc Gadget-3} \citep{Springel2005}.
We adopt its meshless finite-mass (MFM) solver for hydrodynamics \citep{Hopkins2015}
which is a variation of the meshless Godunov method \citep{Gaburov2011}.
Gravity is calculated using the Barnes–Hut method (``treecode'').

\subsection{The ISM model}
In this section,
we summarize the physical processes in the ISM
in our simulations
excluding dust evolution, which will be described in the Sec.~\ref{sec:dustevol}.
The methods and implementations are largely 
based on \citet{Hu2016, Hu2017, Hu2021a}
where more details can be found.

\subsubsection{Time-dependent cooling and chemistry}

We adopt a time-dependent chemistry network developed in 
\citet{Glover2007} and \citet{Glover2012a}
that is widely used in ISM simulations.
The abundances of H$_2$, H$^+$, {\hi}, and the free electron fraction are integrated
based on the chemistry reactions in the network.
The hydrogen network includes 
H$_2$ formation on dust,
H$_2$ destruction by photodissociation, collisional dissociation and cosmic ray ionization,
and recombination in the gas phase and on dust grains.
Individual cooling and heating processes are calculated based on 
the time-dependent chemical abundances,
which includes
cooling from fine structure metal lines,
molecular lines,
Lyman alpha, 
H$_2$ collisional dissociation,
collisional ionization of H, and recombination of H$^+$ in the gas phase and on grains. 
Heating includes photoelectric effect, 
cosmic ray ionization, 
H$_2$ photodissociation, 
UV pumping of H$_2$ and the formation of H$_2$.
Following \citet{Clark2012},
shielding against FUV radiation uses
the {\sc HEALPix} algorithm \citep{Gorski2011}
in combination with the ``treecode'' approximation
to integrate the relevant column densities along 12 sightlines up to a pre-defined radius of 100 pc.
Self-shielding of H$_2$ follows the fitting formula from \citet{Draine1996}\footnote{We do not adopt the improved fitting forumla from \citet{WolcottGreen2011} as most of the H$_2$ forms in low temperatures where the fitting formula from \citet{Draine1996} is a reasonable approximation.
using the H$_2$ column densities obtained by the {\sc HEALPix} method
and the Doppler factor $b$ from gas temperature.
We neglect turbulent broadening that could reduce self-shielding.
However, the Doppler factor $b$ mainly affects self-shielding where the H$_2$ column density is less than $10^{16}~{\rm cm^{-2}}$,
which is a very thin layer we do not resolve.
We adopt a constant cosmic ray ionization rate of {HI} $\zeta_{\rm CR} = 10^{-17} {\rm s^{-1}}$,
an order of magnitude lower than that observed in the Milky Way \citep{Indriolo2015}
to account for the low star formaton rate in our simulated galaxy.
The FUV radiation field is directly calculated from star particles (see Section~\ref{sec:vG0}).
}

\subsubsection{Star formation}

We adopt the stochastic star formation recipe commonly used in the field of galaxy formation.
A gas particle eligible for star formation is converted into a star particle of the same mass
on a timescale of $t_{\rm ff} / \epsilon_{\rm sf}$ stochastically,
where 
$\epsilon_{\rm sf}$ is the star formation efficiency
and 
$t_{\rm ff} = \sqrt{ 3\pi /  (32 G \rho) }$ is the gas free-fall time
where
$G$ is the gravitational constant and
$\rho$ is the gas density.
We adopt $\epsilon_{\rm sf} = 0.5$ in this work.
Such a high efficiency is justified by our resolution 
as we can follow the gravitational collapse down to the scales of individual molecular cores.
Gas is eligible for star formation
when its local 
Jeans mass $M_{\rm J} = (\pi^{2.5} c_s^3)/(6 G^{1.5} \rho^{0.5})$,
drops below the kernel mass $M_{\rm ker} = N_{\rm ngb} m_{\rm g}$,
where $c_s$ is the sound speed, 
$m_{\rm g}$ is the gas particle mass,
and $N_{\rm ngb} = 32$ is the number of neighboring particles in a kernel.

\subsubsection{Sampling individual stars from an IMF}
Massive stars (initial mass $> 8\ {\rm M_\odot}$) inject energy and momentum into their surrounding gas
commonly termed as ``stellar feedback''. 
At our resolution of $1~{\rm M_\odot}$ per star particle,
it is unphysical to assume that each particle represents 
a star cluster with a fully sampled stellar initial mass function (IMF),
as is commonly the case in cosmological simulations with much coarser resolutions.
Following \citet{Hu2021a}, 
we adopt the technique of ``importance sampling'' 
to stochastically sample stellar masses from a Kroupa IMF \citep{Kroupa2001}.
The sampled stellar masses
are used to determine the stellar lifetime \citep{Ekstroem2012} and 
UV luminosity from the BaSeL stellar library \citep{Lejeune1997, Lejeune1998}
This does not affect the dynamics,
as the gravitational masses of star particles ($m_*$) remain unchanged.

\subsubsection{Stellar feedback}

We include stellar feedback from supernovae (SNe) and photoionization.
SN feedback is done by injecting thermal energy of $10^{51}$~erg per SN 
into its nearest $N_{\rm ngb}$ gas particles in a kernel-weighted fashion.
As our resolution is able to resolve the Sedov-Taylor phase in each SN event,
a simple thermal feedback is able to achieve numerical convergence \citep{Hu2019}.
Feedback from photoionization follows \citet{Hu2017},
where
each massive star searches for its ionization front iteratively by balancing recombination and photoionization 
and heats up the interior gas to $10^4$ K.
The hydrogen is assumed to be fully ionized in the HII regions where chemistry is turned off.
This approach reproduces the dynamics of an expanding {\hii} region 
predicted by radiative transfer codes in a uniform medium.
More importantly,
it captures the correct behaviors in overlapping {\hii} regions
where a naive Str\"{o}mgren-sphere method would suffer from the numerical artifact of double-counting.

\subsubsection{Spatially variable FUV radiation}\label{sec:vG0}

Following \citet{Hu2017},
the unattenuated FUV radiation field is both spatially and temporally variable
and is calculated directly from the star particles.
For a given gas particle,
every star particle contributes a radiation flux of $L_{\rm FUV} / (4\pi r^2)$
where $L_{\rm FUV}$ is the FUV luminosity based on the sampled stellar mass
and $r$ is the distance between the gas and star particle.
The summation is over all star particles
and is done via the ``treecode'' approximation to 
avoid the $O(n^2)$ operation and
speed up the calculation.
The FUV radiation affects both the thermal balance via photoelectric heating
and the chemistry via photodissociation.

\subsection{Dust evolution model}\label{sec:dustevol}

We adopt the ``one-fluid'' approach where
dust is assumed to be spatially coupled with the gas.
This is justified as dust is expected to be charged and gyrates around the magnetic fields in the ISM 
with a small gyro-radius.
Each gas particle is associated with a dust mass 
$m_{\rm d} = m_{\rm d}({\rm Sil}) + m_{\rm d}({\rm C})$,
where $m_{\rm d}({\rm Sil})$ and $m_{\rm d}({\rm C})$
are, respectively, the masses of silicate dust and carbonaceous dust,
which evolve separately in the simulations
due to the production and destruction processes
as we describe below.
Dust grains are assumed to be spherical 
with a radius of $a$ and a material density of $s_d$,
which leads to a grain mass of $m_{\rm gr} = (4 \pi a^3/3) s_d$.
The formation or destruction rate of dust can be expressed as
\begin{equation}
	\frac{ {\rm d} m_{\rm d}}{ {\rm d} t} = N_{\rm gr} \frac{ {\rm d} m_{\rm gr}}{ {\rm d} t} = 3 \frac{\dot{a}}{a} m_{\rm d}
\end{equation}
where $N_{\rm gr} = m_{\rm d} / m_{\rm gr}$ is number of dust grains in a gas cell.
Time integration is done via sub-cycling in order to resolve the timescales of dust dynamics and sputtering 
that can be orders of magnitude smaller than the hydrodynamical timesteps.


We include physical processes that directly modify $m_{\rm d}$:
sputtering, dust growth, and dust formation in AGB ejecta.
Processes that modify the grain size while keeping $m_{\rm d}$ fixed,
such as shattering and coagulation,
are not included.
This is because we are only interested in the evolution of dust mass rather than 
other dust properties such as the extinction law.
We adopt a fixed grain-size distribution that follows \citet{Mathis1977} (the ``MRN'' distribution).


\subsubsection{Sputtering}

In shocks or in hot gas,
dust can be destroyed via sputtering which returns metals locked up in dust grains back to the ISM.
We adopt the sputtering model in \citet{Hu2019a}
that includes thermal and nonthermal sputtering,
which we briefly summarize as follow.
The dust destruction rate due to sputtering can be expressed as 
\begin{equation}
	\frac{ {\rm d} m_{\rm d}}{ {\rm d} t} \Big|_{\rm sput}= - \frac{m_{\rm d}}{t_{\rm sput}}
\end{equation}
where 
\begin{eqnarray}
	t_{\rm sput} 
	&\equiv& \frac{a}{3 n Y_{\rm tot}} \nonumber\\
	&=& 10~{\rm kyr}\ \Big(\frac{a}{0.03 \mu m}\Big) \Big(\frac{n}{{\rm cm}^{-3}}\Big)^{-1} 
	\Big(\frac{10^{6} ~Y_{\rm tot}}{\rm \mu m \ yr^{-1}cm^3}\Big)^{-1},\nonumber\\ \label{eq:t_sput}
\end{eqnarray}
where 
$n$ is the hydrogen number density and
$Y_{\rm tot}$ is the erosion rate
that includes thermal and nonthermal sputtering,
which we adopt from \citet{Nozawa2006}.
The thermal erosion rate is a function of gas temperature
while the nonthermal erosion rate is a function of the relative bulk velocity between dust and gas,
which we obtain by 
integrating the equation of motion for dust 
accounting for
direct collision, plasma drag, and betatron acceleration.

\subsubsection{Dust growth}

Dust can grow
in the cold gas
when 
gas-phase metals
interact with dust and stick onto the surfaces of dust grains,
which can be viewed as the reverse process of sputtering.
The exact mechanism is still poorly understood,
though its feasibility has been supported by laboratory experiments 
\citep{Krasnokutski2014, Henning2018, Rouille2020}.
The dust production rate due to dust growth can be expressed as
\begin{equation}\label{eq:ODEgrow}
	\frac{ {\rm d} m_{\rm d}}{ {\rm d} t} \Big|_{\rm grow} = (1 - f) \frac{m_{\rm d}}{t_{\rm grow}}
\end{equation}
where 
$t_{\rm grow}$ is the
dust growth timescale (see below)
and 
$f$ is the fraction of metals locked in dust grains.
For element $A$ (where $A = $ Si or C),
this can be written as
\begin{equation}
	f_{\rm A} = \frac{m_{\rm A,d}}{m_{\rm A,tot}} = \frac{ m_{\rm d}({\rm A}) \xi_{\rm A} }{ m_{\rm g} X_{\rm A} Z^\prime }, 
\end{equation}
where
$m_{\rm A,d}$ is the mass of element $A$ in the dust phase,
$m_{\rm A,tot}$ is the total mass of element $A$ (dust + gas),
$X_{\rm A}$ is the solar abundance of element $A$,
and $\xi_{\rm A}$ is the mass fraction of element $A$ in the assumed grain material.
For carbonaceous dust, $\xi_{\rm C} = 1$.
For silicate dust,
we adopt ${\rm MgFeSiO_4}$ as the grain material
which leads to $\xi_{\rm Si} = 0.165$.
The dust growth timescale takes the following form 
\begin{eqnarray}\label{eq:t_grow}
	t_{\rm grow} 
	&=& 1.5~\text{Gyr}~\Big(\frac{ n }{ \text{cm}^{-3} }\Big)^{-1} \Big( \frac{T}{100\text{K}} \Big)^{-0.5}  \nonumber\\
	&& \Big( \frac{a_3}{0.03 \upmu\text{m}} \Big) (Z^{\prime} \alpha_s D_{\rm eff})^{-1}.
\end{eqnarray}
Here,
$\alpha_s$ is the sticking coefficient,
$a_3 \equiv \langle a^3 \rangle /  \langle a^2 \rangle$ is the average grain size
where the bracket $\langle \ldots \rangle$ refers to integration over the grain size distribution,
and 
$D_{\rm eff} \equiv  \langle a^2 D(a) \rangle /  \langle a^2 \rangle$
is the enhancement factor $D(a)$ due to Coulomb focusing \citep{Weingartner1999}
weighted by the surface area of grains.
The sticking coefficient encompasses
the complex physical and chemical processes on the surfaces of grains
which is still uncertain (see, e.g., \citealp{Zhukovska2018} for a discussion).
To first order, 
it is expected to be close to unity at low temperatures
and decrease substantially at high temperatures.
We follow \citet{Zhukovska2016} and
assume that $\alpha_s = 1$ at $T < 300$~K while
$\alpha_s = 0$ at $T \geq 300$~K.
The area-weighted enhancement factor $D_{\rm eff}$ is also uncertain
and is expected to depend on a number of properties
such as density, temperature, free electron fraction, 
grain size distribution, and grain charge.
\citet{Weingartner1999} found that Coulomb focusing
shortens the growth timescale by more than an order of magnitude.
However,
\citet{Priestley2021} found a much weaker effect if the evolution of grain size is taken into account.
We adopt $D_{\rm eff} = 10$ for simplicity,
but 
the uncertainties in both $\alpha_s$ and $D_{\rm eff}$ are potential caveats of our model.


\subsubsection{Dust production from AGB stars}
We adopt the mass-dependent dust yields 
from \citet{Zhukovska2008} at $Z^\prime = 0.1$
based on the individual stellar masses sampled from the IMF
to account for dust produced in the ejecta of AGB stars.
The produced dust mass is injected in the neighboring gas particles
in a kernel-weighted fashion.
We do so only for carbonaceous dust
as the silicate dust yields at this metallicity is essentially zero.
Dust produced from AGB stars 
is expected to play a sub-dominant effect on the spatial variation of the DGR
as AGB stars are more uniformly distributed in the ISM
compared to sputtering in SN shocks and dust growth in dense clouds
that are highly clustered.


\subsection{Chemistry network in post-processing}

In order to accurately model the transitions of C$^+$/{\ci}/CO (the ``carbon cycle''),
a detailed carbon chemistry is required, 
which is much more complicated and computationally costly
to solve compared to the hydrogen chemistry.
We therefore post-process the simulation snapshots 
using {\sc AstroChemistry}\footnote{Available at \url{https://github.com/huchiayu/AstroChemistry.jl}.},
a chemistry network code
developed in \citet{Hu2021a}.
It consists of 31 species:
H, H$^-$, H$_2$, H$^+$, H$_2^+$, H$_3^+$, e$^-$, He, He$^+$, HeH$^+$,
C, C$^+$, CO, HCO$^+$,
O, O$^+$, OH, OH$^+$, H$_2$O$^+$, H$_3$O$^+$, H$_2$O,
O$_2$, CO$^+$, O$_2^+$,
CH$_2$, CH$_2^+$, CH, CH$^+$, CH$_3^+$,
Si$^+$ and Si.
All chemical reactions in the UMIST database 
\citep{McElroy2013} that exclusively involve the above-mentioned species 
are included in the network,
which leads to 286 reactions in total.
The time-dependent abundances of H$_2$ and H$^+$ in the simulations are taken as input parameters when solving the network.
This is crucial as H$_2$ can be out of steady state significantly, especially at low metallicity.
The free electron abundance is calculated from charge conservation,
which is dominated by H$^+$ in the diffuse ISM and by C$^+$ in the dense gas.
Both dust shielding and gas shielding (including H$_2$ and CO) for CO are included
using a {\sc HealPIX} method similar to the time-dependent chemistry\footnote{
We do so iteratively with two iterations which has been shown to be sufficient in \citet{Hu2021a}.}.
This approach has been applied to ISM-patch simulations
that 
successfully reproduced the observed relationship between 
the column densities of CO and H$_2$ in Galactic clouds \citep{Hu2021a}
as well as
the Milky Way CO-to-H$_2$ conversion factor ($X_{\rm CO}$) \citep{Hu2022a}.
We do not include the effect of dust depletion in post-processing and instead assume 
constant gas-phase metal abundances. 
The significance of depletion depends on the competition between dust growth 
and CO formation that requires time-dependent carbon chemistry, 
which is an interesting topic for future work.

\citet{Hu2022a} found that {\coone} is almost always optically thin 
in their simulated ISM
at $Z^\prime = 0.1$.
This is primarily a direct consequence of the low CO abundance.
Furthermore,
{\coone} can remain optically thin even at the highest column densities 
as the population of CO is distributed over more excited levels.
Assuming optically thin conditions,
the CO luminosity from each gas particle can be expressed by
\begin{eqnarray}
	l_{\rm CO} = 0.5 \lambda_{10}^3 T_{10} A_{10} n_{\rm CO} f_1 \overline{V} 
\end{eqnarray}
where 
$n_{\rm CO}$ is the CO number density,
$\overline{V} = m_{\rm g} / \rho$ is the volume of the gas particle,
and
$\lambda_{10} = 0.26$~cm,
$T_{10} = 5.53$~K,
$A_{10} = 7.2\times 10^{-8}~{\rm s}^{-1}$
are the wavelength, energy level, and the Einstein A coefficient for the {\coone} line, respectively.
$f_1(T_{\rm ex}) = 3\exp(-5.53 /T_{\rm ex}) / \sqrt{1 + (T_{\rm ex}/2.77)^2}$ 
is the fraction of CO in the level $J=1$ \citep{Draine2011}.
Note that $f_1$ only varies within a factor of 2 in the range of $3 < T_{\rm ex}/{\rm K} < 30$ where most CO exists.

\subsection{Simulation setup}

The initial conditions 
consist of a rotating disk galaxy embedded in a dark matter halo
with properties resembling the WLM galaxy,
generated by the {\sc MakeDiskGalaxy} code \citep{Springel2005}.
The halo has a virial radius $R_{\rm vir}$ = 45 kpc and a virial mass $M_{\rm vir}$ = $10^{10}~{\rm M}_\odot$,
and it follows a Hernquist profile \citep{Hernquist1990}
matching an NFW \citep{Navarro1997} profile at small radii
with the concentration parameter $c$ = 15 and the spin parameter $\lambda$ = 0.035.
The baryonic mass fraction is 0.8\%,
with a stellar disk of $10^7~{\rm M}_\odot$ and a gaseous disk of $7\times 10^7~{\rm M}_\odot$,
both following an exponential profile with a scale-length of 1 kpc.
The central gas surface density is $\Sigma_{\rm gas} \sim 10~{\rm M_\odot~pc^{-2}}$.
The stellar disk follows an exponential vertical profile with a scale-height of 1 kpc.
The vertical density profile of the gaseous disk is set up to maintain hydrostatic equilibrium.
The initial gas temperature is set to be $10^4$ K.
The particle mass for gas, stars, and dark matter
are $m_{\rm g} = 1~{\rm M_\odot}$, 
$m_{*} = 1~{\rm M_\odot}$, 
and
$m_{\rm dm} = 10^3~{\rm M_\odot}$, 
respectively.
The corresponding spatial resolution for gas is $\sim 0.2$~pc,
defined as the minimum kernel radius where the Jeans length can be resolved \citep{Hu2021a}.
Such a high resolution is needed in order to resolve the dense and compact cores 
where CO is observed in the WLM galaxy.
The gravitational softening length is $0.2$~pc for gas and 100~pc for the dark matter.


The metallicity is $Z^\prime \equiv Z / Z_\odot = 0.1$ throughout the simulations.
The initial dust-to-gas ratio is set to be 1\% of the Milky Way value,
$Z^\prime_{\rm d} \equiv  Z_{\rm d} / Z_{\rm d,MW} = 0.01$,
motivated by observations of low-metallicity galaxies \citep{RemyRuyer2014}.
We adopt
the Milky Way dust abundances
as
$Z_{\rm d,MW} ({\rm C}) = 1.9\times 10^{-3}$ and
$Z_{\rm d,MW} ({\rm Sil})= 3.5\times 10^{-3}$ \citep{Dwek2005}
corresponding to a carbonaceous-to-silicate ratio $\sim 0.54$
and 
a total dust-to-gas ratio of $Z_{\rm d,MW} = 5.4\times 10^{-3}$.

Galaxy scale simulations with a setup like ours are known to undergo 
an artificial burst of star formation during the initial collapse,
which in turn leads to 
overly-energetic SN feedback that blows out the entire gaseous disk,
substantially reducing the gas surface density.
Following \citet{Hu2022},
we minimize this artifact by first running a simulation without dust evolution for 100 Myr 
and with the SN delay time set to zero.
This reduces the dynamical impact of feedback due to suppressed SN clustering.
The simulation snapshot at the end of this ``pre-simulation'' is used as the new initial conditions
with a relaxed configuration.

We run three simulations:
(1) a fiducial model with fully coupled chemistry and dust evolution,
(2) a model without dust evolution (i.e., $Z^\prime_{\rm d} = 0.01$ throughout the simulation),
and 
(3) a model with dust evolution where dust growth is switched off.
Each simulation is run for 0.5 Gyr.


\begin{figure*}
	\centering
	\includegraphics[trim = 0cm 0.0cm 0cm 1.5cm, clip, width=0.99\linewidth]{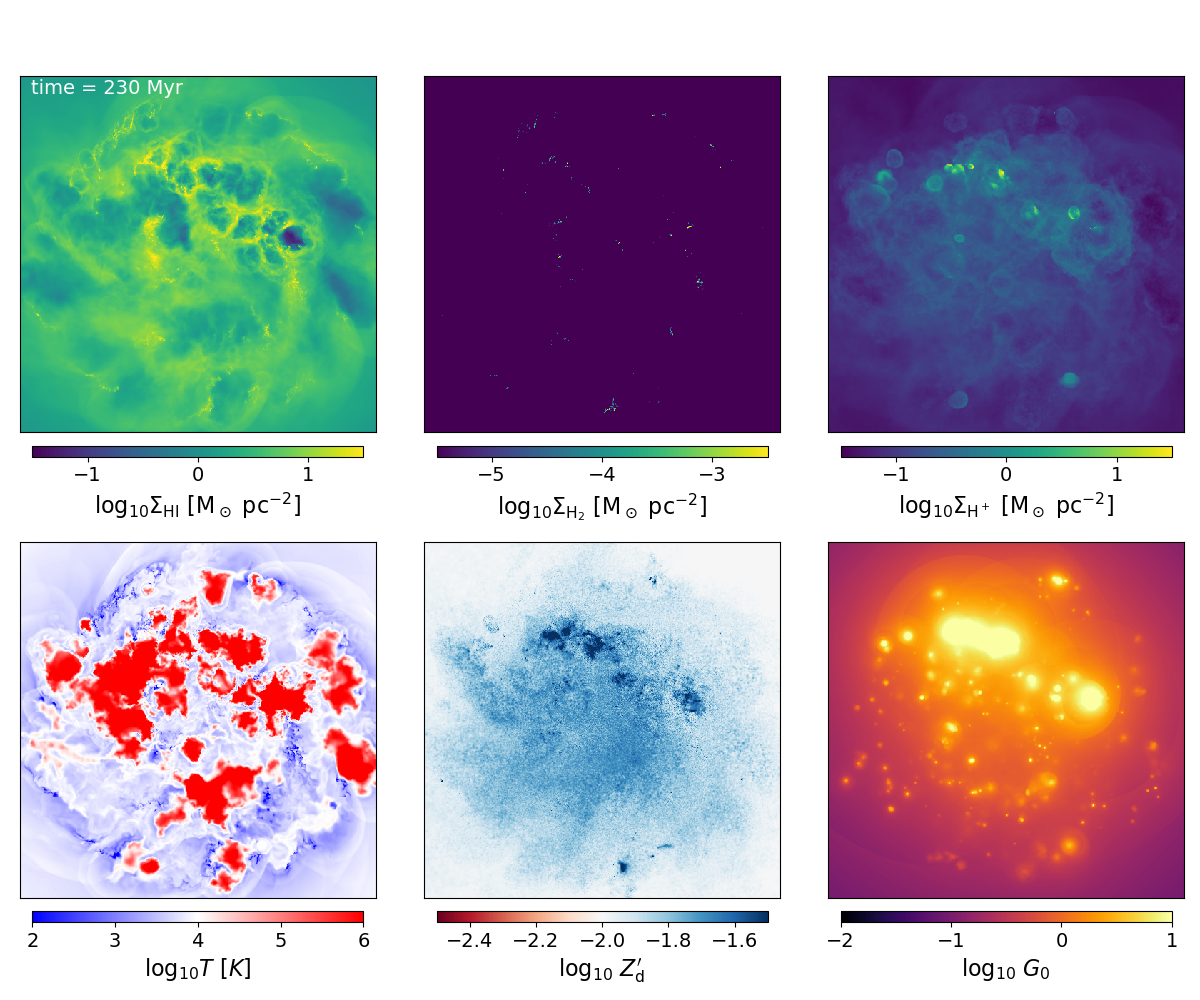}
	\caption{Face-on images of 
		the surface densities of {\hi}, H$_2$, and H$^+$,
		gas temperature ($T$),
		projected DGR ($Z^{\rm \prime proj}_{\rm d})$,
		and FUV radiation field ($G_0$, in units of the \citealp{Habing1968} field)
		at simulation time $t = 230$~Myr.
	}
	\label{fig:maps_230}
\end{figure*}

\section{Results} \label{sec:results}

\subsection{Overview}

Fig.~\ref{fig:maps_230} shows the face-on images of 
the surface densities of {\hi}, H$_2$, and H$^+$,
gas temperature,
projected DGR,
and FUV radiation field
at simulation time $t = 230$~Myr.
The ISM has a complex structure,
with dense clouds where star formation occurs and holes driven by stellar feedback (``SN bubbles'').
The majority of gas is in the form of {\hi} 
while H$^+$ traces the young massive stars
photoionizing the ambient gas (the {\hii} regions).
The FUV radiation also traces young stars but it is more spatially extended.
The H$_2$ abundance is extremely low everywhere besides the densest part of clouds.
These pc-scale dense cores are also where CO exists (cf.~Fig.~\ref{fig:nH_xi}).
Most of the gas is in the warm phase with a temperature of $T\sim 10^4$~K,
while the hot gas ($T\sim 10^6$~K) is found in the interior of the SN bubbles.
Cold gas with $T\sim 100$~K only exists in dense and compact clouds.


The projected DGR image demonstrates that the DGR is not spatially uniform.
The DGR is elevated in dense gas where dust growth is most efficient.
Interestingly,
the DGR is not suppressed 
in the interior of SN bubbles
where dust is expected to be destroyed via sputtering.
Instead,
the DGR seems to be elevated inside the SN bubbles.
More quantitative analysis will be provided in Sec.~\ref{sec:dist_dgr}.

\subsection{Effect on ISM chemistry}

\begin{figure*}
	\centering
	\includegraphics[trim = 1.8cm 0cm 1.8cm 0cm, clip, width=0.95\linewidth]{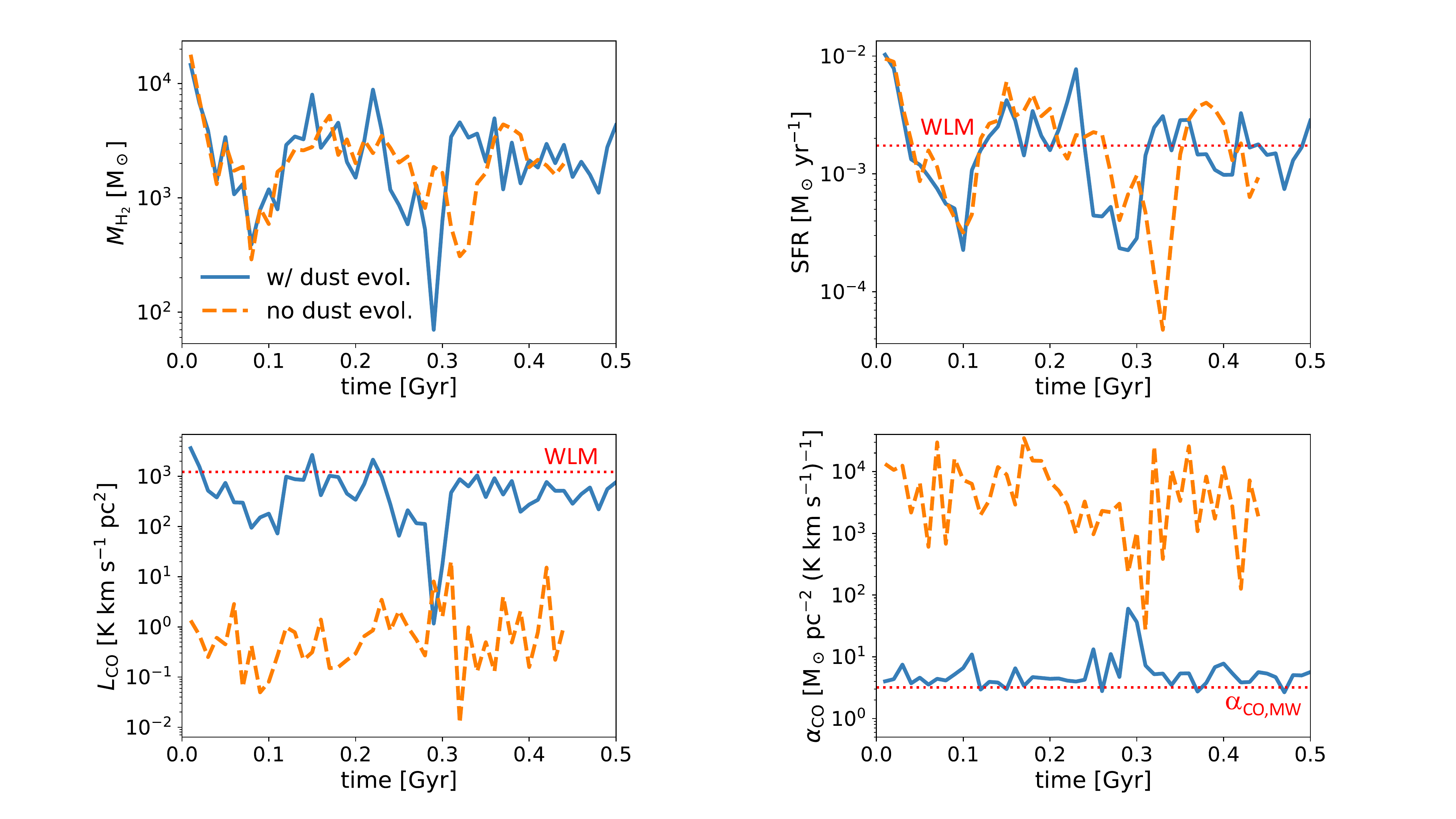}
	\caption{
		Time evolution of the following global properties integrated over the simulated galaxy:
		the H$_2$ mass ($M_{\rm H_2}$, top left), the star formation rate (SFR, top right),
		the luminosity of the {\coone} emission ($L_{\rm CO}$, bottom left),
		and the CO-to-H$_2$ conversion factor 
		($\alpha_{\rm CO} \equiv M_{\rm H_2} / L_{\rm CO}$, bottom right).
		The solid blue lines represent our fiducial run including dust evolution 
		while the dashed orange lines represent a controlled run without dust evolution.
		The red dotted lines indicate the observed $L_{\rm CO}$ and SFR in the WLM galaxy
		as well as the $\alpha_{\rm CO}$ in the Milky Way.
		Dust evolution strongly enhances $L_{\rm CO}$, but it has little effect on $M_{\rm H_2}$ and SFR.
	}
	\label{fig:time_H2COSFR}
\end{figure*}

Fig.~\ref{fig:time_H2COSFR} shows the following global quantities of the simulated galaxy as a function of time:
the H$_2$ mass ($M_{\rm H_2}$, top left), the star formation rate (SFR, top right),
the luminosity of the {\coone} emission ($L_{\rm CO}$, bottom left),
and the CO-to-H$_2$ conversion factor ($\alpha_{\rm CO} \equiv M_{\rm H_2} / L_{\rm CO}$, bottom right).
Two simulations are shown, one with dust evolution (solid blue lines, our fiducial model) and one without dust evolution
where the DGR is constant everywhere (dashed orange lines).
The WLM galaxy has an observed CO luminosity of 
$1229~{\rm K~km~s^{-1}~pc^2}$ \citep{Rubio2015}
and an observed SFR of 
$1.74\times 10^{-3}~{\rm M_\odot~yr^{-1}}$  \citep{2010AJ....139..447H},
as indicated by the red dotted lines.
The CO-to-H$_2$ conversion factor in the Milky Way
$\alpha_{\rm CO,MW} = 3.2~{\rm M_\odot~pc^{-2}~(K\ km~s^{-1})^{-1}}$ (excluding helium)
is also overplotted as the red dotted line.
Table~\ref{tab:sumstats} summarizes the median values of these global quantities.

\begin{figure*}
	\centering
	\includegraphics[trim = 0cm 0.2cm 0cm 0cm, clip, width=0.75\linewidth]{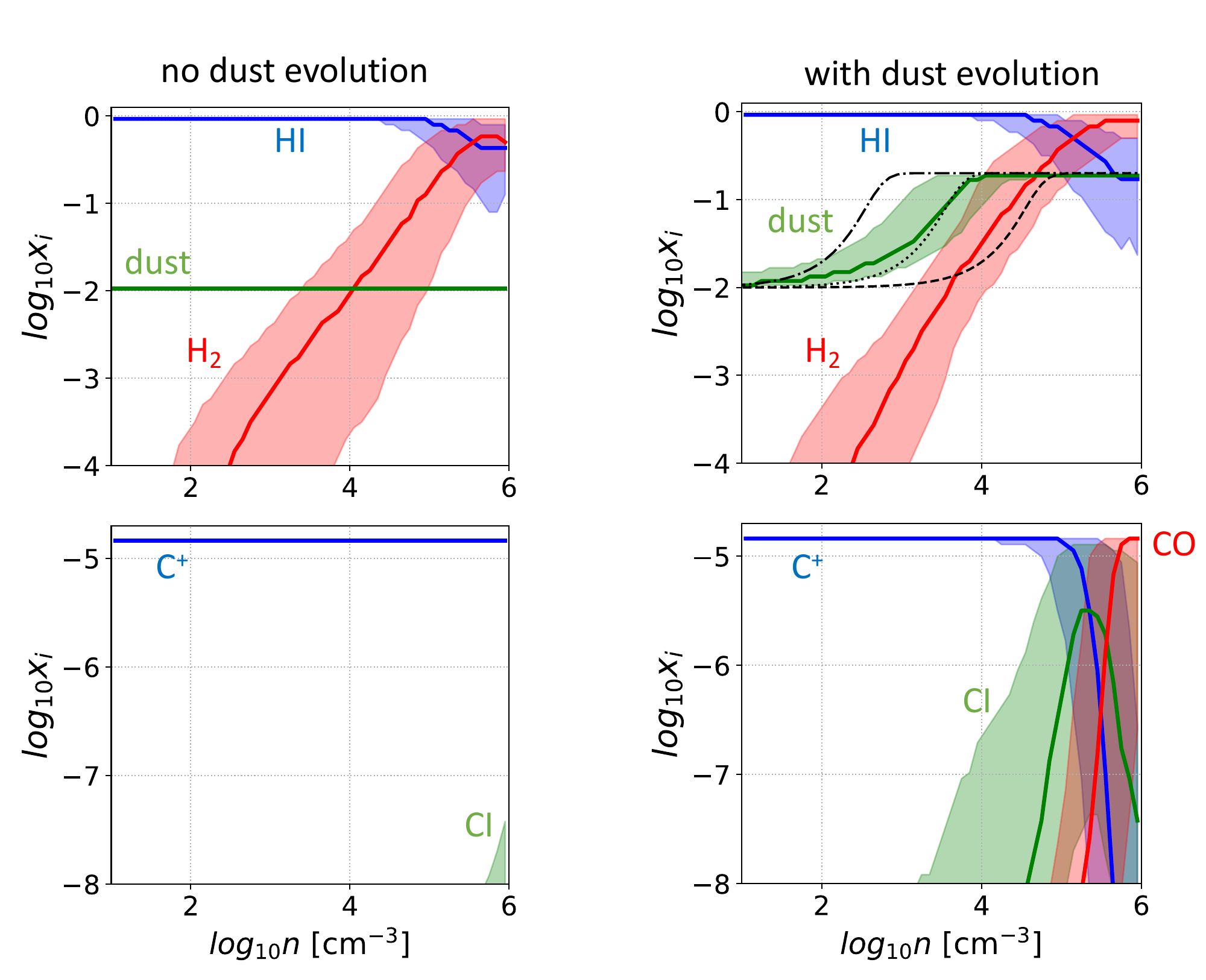}
	\caption{
		Top panels: 
		the DGR normalized to the Milky Way value $Z^\prime_{\rm d}$ (green) and chemical abundances of {\hi} (blue) and H$_2$ (red) as a function of the hydrogen number density $n$.
		Bottom panels:
		chemical abundances of C$^+$ (blue), {\ci} (green), and CO (red) as a function of $n$.		
		The right and left panels are the runs with and without dust evolution, respectively.
		The solid lines show the median value in a given $n$ bin
		while the shaded area brackets the 16 and 84 percentiles.
		The black lines in the upper right panel 
		indicate the analytic solution in
		Eq.~\ref{eq:ODEgrow} with the dynamical time 
		$t_{\rm dyn} = $ 0.1 (dashed), 1 (dotted), and 10 Myr (dash-dotted), respectively.
		Without dust evolution,
		CO is completely photo-dissociated due to insufficient shielding while H$_2$ is almost unaffected.
	}
	\label{fig:nH_xi}
\end{figure*}

Our fiducial model successfully reproduces the observed
$L_{\rm CO}$ and SFR 
in the WLM galaxy.
On the other hand,
$M_{\rm H_2}$ is extremely low and 
is only a fraction $\sim 10^{-4}$ of the ISM mass.
This is due to the long H$_2$ formation time compared to the dynamical time in the highly turbulent ISM
and is consistent with previous simulations of isolated dwarf galaxies \citep{Hu2016, Hu2017, Whitworth2022}.
The low $M_{\rm H_2}$ leads to 
a conversion factor close to $\alpha_{\rm CO,MW}$ despite the low metallicity.
Dust evolution has a substantial impact on the CO luminosity.
Without dust evolution,
$L_{\rm CO}$ is about three orders of magnitude lower than the observed value.
On the other hand,
$M_{\rm H_2}$ and the SFR are both insensitive to dust evolution.

Our low $\alpha_{\rm CO}$ may seem to be in conflict with 
\citet{Hu2022a} where the kpc-scale $\alpha_{\rm CO}$ was found to scale with $Z^{\prime -0.71}$,
implying $\alpha_{\rm CO} \sim 5 \alpha_{\rm CO,MW}$ at $Z^\prime = 0.1$.
This is because we adopt $Z_{\rm d}^\prime = 0.01$ in this work 
(motivated by the observed super-linear metallicity--DGR relation),
which is ten times lower than what \citet{Hu2022a} assumed.
As a result,
the molecular mass fraction $F_{\rm H_2}$ is also about ten times lower,
which explains the difference in $\alpha_{\rm CO}$.


\begin{deluxetable*}{ccccccc}
	\tablecaption{Median values of global properties over $100 < t  < 500$~Myr.
		\label{tab:sumstats}
	}
	\tablewidth{0pt}
	\tablehead{
		\colhead{model} &
		\colhead{${\rm SFR}$}  &
		\colhead{$M_{\rm H_2}$ }  &
		\colhead{$M_{\rm CO}$}   &
		\colhead{$L_{\rm CO}$}   &
		\colhead{$\alpha_{\rm CO}$}   &
		\colhead{${\rm SFR} / L_{\rm CO}$}  \\ 
		\colhead{} &
		\colhead{(1)} &
		\colhead{(2)} &
		\colhead{(3)} &
		\colhead{(4)} &
		\colhead{(5)} &
		\colhead{(6)} 
	}
	\startdata
	w/ dust evolution   &$1.59\times 10^{-3}$  &  2435  & $1.64\times 10^{-1}$ &  469  &  4.63	  &  $3.41\times 10^{-6}$   	 \\
	no dust evolution   &$2.02\times 10^{-3}$  &  2095  & $1.97\times 10^{-4}$ &  0.72 &  3156	  &  $2.19\times 10^{-3}$   	 \\
	\enddata
	\tablecomments{
		(1) Star formation rate [${\rm M_\odot~yr^{-1}}$].
		(2) Total H$_2$ mass [${\rm M_\odot}$]
		(3) Total CO mass [${\rm M_\odot}$]
		(4) Total {\coone} luminosity [${\rm K~km~s^{-1}~pc^2}$]
		(5) CO-to-H$_2$ conversion factor $\alpha_{\rm CO} \equiv M_{\rm H_2} / L_{\rm CO}$ [$\rm M_\odot~pc^{-2}~(K~km~s^{-1})^{-1}$]
		(6) Ratio of ${\rm SFR} / L_{\rm CO}$ $[\rm M_\odot~yr^{-1}~pc^{-2}~(K~km~s^{-1})^{-1}]$
	}
\end{deluxetable*}

We now take a closer look at the local chemical abundances and DGR as a function of hydrogen number density $n$
as shown in Fig.~\ref{fig:nH_xi}.
The top panels show the DGR and abundances of H$_2$ and {\hi} 
while the bottom panels show the abundances of C$^+$, {\ci}, and CO.
The right and left panels are models with and without dust evolution, respectively.

Dust growth enhances
$Z^\prime_{\rm d}$ only at high enough densities where $n > 10^3~{\rm cm}^{-3}$.
This is a result of the short dynamical time in the ISM ($t_{\rm dyn}$)
which limits the available time
for dust growth to operate before the dense clouds are destroyed.
Indeed,
the dust growth rate in Eq.~\ref{eq:ODEgrow} in a static medium 
has the following analytic solution \citep{Zhukovska2008}:
\begin{equation}
	f(t) = f(0) \frac{\exp(t / t_{\rm grow})}{1 - f(0) + f(0) \exp(t / t_{\rm grow}) },
\end{equation}
where $f(0)$ is the initial dust depletion fraction
and $t_{\rm grow}$ is given in Eq.~\ref{eq:t_grow} which is density-dependent.
For a given $t_{\rm dyn}$,
we can therefore construct the analytic solution as a function of $n$,
as shown in black lines in the upper right panel in Fig.~\ref{fig:nH_xi}
for $t_{\rm dyn} = $ 0.1 (dashed), 1 (dotted), and 10~Myr (dash-dotted), respectively.
The median $Z^\prime_{\rm d}$ from our fiducial simulation 
can be reproduced remarkably well with $t_{\rm dyn} = 1$~Myr.
This Myr-scale dynamical time in the ISM is consistent with \citet{Hu2021a}.

The fact that dust growth only operates at high densities ($n > 10^3~{\rm cm}^{-3}$)
means that it only modestly increases the H$_2$ formation rate on dust grains.
Meanwhile,
dust growth can also enhance radiation shielding in dense gas.
However,
this does not affect the H$_2$ abundance very much either
as (1) H$_2$ can self-shield against the FUV radiation and 
(2) the limiting factor for H$_2$ is the available time for it to form while shielding only plays a secondary role \citep{Glover2011, Hu2021a}.
Indeed, 
the H$_2$ abundance is only notably enhanced at $n > 10^4~{\rm cm}^{-3}$ with dust evolution.
Similarly,
the SFR is insensitive to dust growth 
because the thermal balance of the ISM is mostly unaffected except for the densest gas.

In contrast,
dust evolution
has a very significant effect on the C$^+$/{\ci}/CO transitions.
Without dust evolution,
both {\ci} and CO are completely destroyed by the FUV radiation.
This is a natural consequence of the adopted low $Z^\prime_{\rm d}$ as motivated by observations.
On the other hand,
with dust evolution,
C$^+$/{\ci}/CO transitions take place at very high densities ($n \gtrsim 10^5~{\rm cm}^{-3}$) 
due to enhanced dust shielding.
This is consistent with the cloud simulations in \citet{Glover2012b}
where they found that the CO luminosity of low-metallicity clouds 
is dominated by emission from gravitationally collapsing dense gas of similar densities.
Note that 
the high-$Z^\prime_{\rm d}$ gas occupies a very small mass fraction in the ISM such that
the global galaxy-integrated $Z^\prime_{\rm d}$ is only $\sim 20\%$ 
higher than in the constant-DGR run.

Due to the long H$_2$ formation time,
gas with $n\sim 100~{\rm cm^{-3}}$,
the typical density for molecular clouds in the Milky Way,
is completely dominated by {\hi}.
In other words,
there is very little CO-dark H$_2$ gas in the ISM as often assumed at low metallicity.
This leads to a galaxy-integrated $\alpha_{\rm CO}$ factor only 50\% higher than the Milky Way value as shown in Fig.~\ref{fig:time_H2COSFR}.

\subsection{Spatial variation of DGR}\label{sec:dist_dgr}

\begin{figure*}
	\centering
	\includegraphics[width=0.99\linewidth]{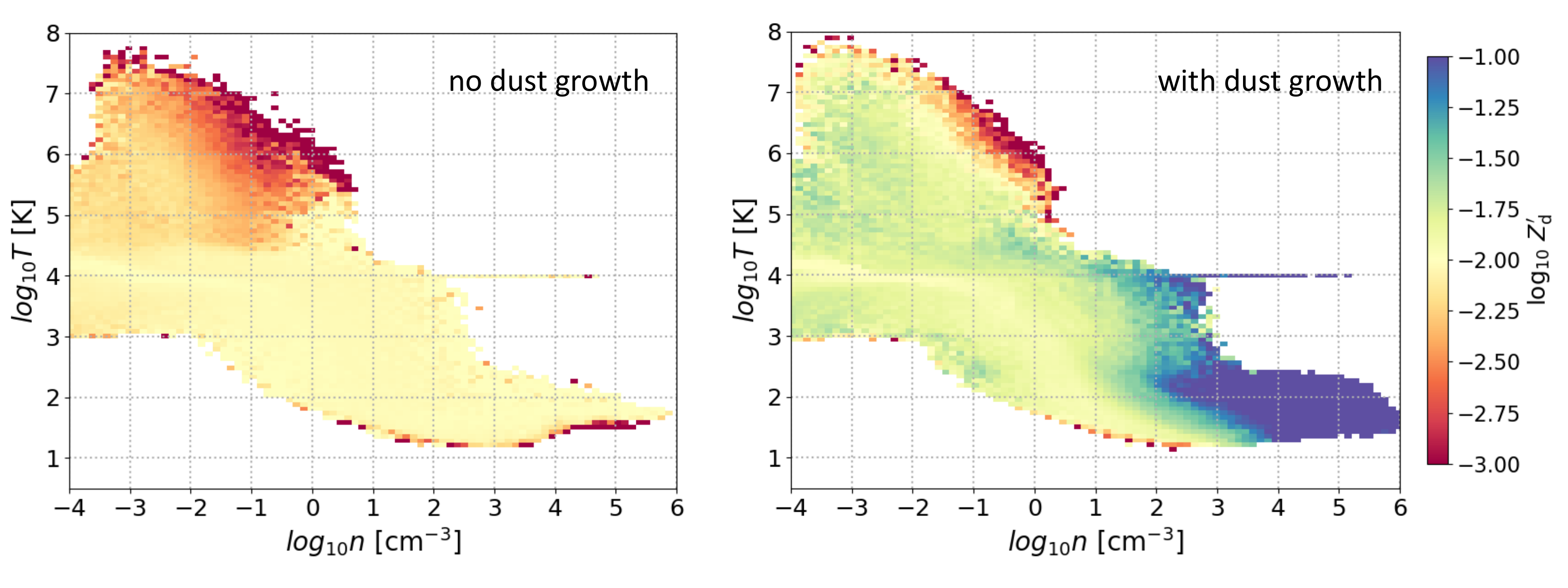}
	\caption{
		The time-averaged phase diagram (density vs. temperature) color-coded by the DGR.
		The right and left panels are for runs with and without dust growth, respectively.
		Dust growth enhances the DGR in dense gas which mixes with the hot gas generated by SN feedback.
	}
	\label{fig:PD_dgr_growth}
\end{figure*}

\begin{figure}
	\centering
	\includegraphics[width=1.0\linewidth]{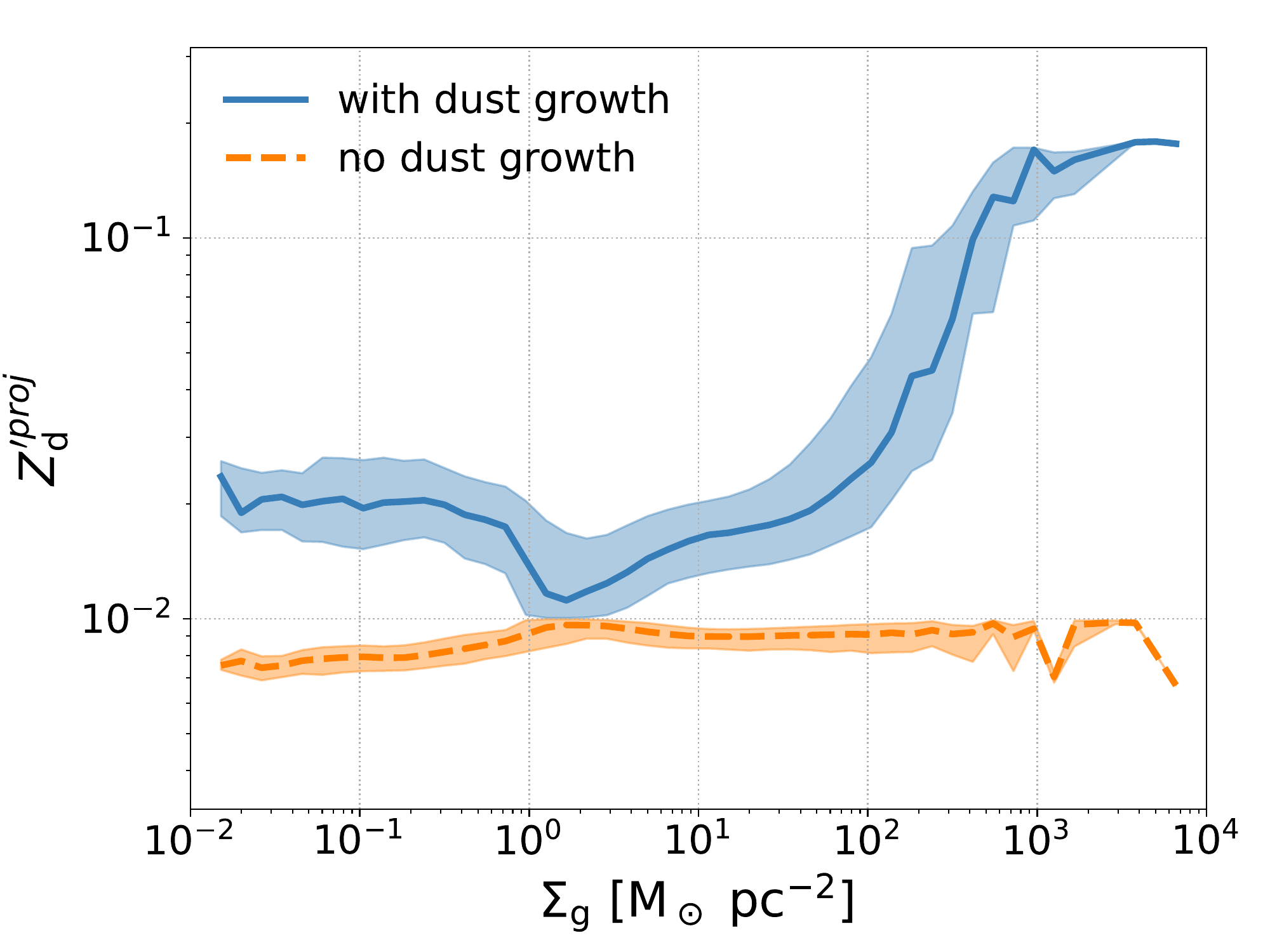}
	\caption{
		The time-averaged projected DGR as a function of the gas surface density with a pixel size of 3~pc.
		The blue solid line shows our fiducial model while 
		the orange dashed line shows the model without dust growth.
		The lines show the median in each bin while the shaded area brackets the 16th and 84th percentiles.
		Dust growth is the main driver of the DGR variation in the ISM.
	}
	\label{fig:Sigma_vs_DGR}
\end{figure}

We now consider how the DGR spatial variation comes about.
We compare our fiducial simulation with the simulation where dust growth is switched off
to investigate the relative importance between dust growth and sputtering.

Fig.~\ref{fig:PD_dgr_growth} shows
the time-averaged phase diagram (density vs.~temperature) color-coded by $Z^\prime_{\rm d}$.
The right and left panels are for runs
with (right panel) and without (left panel) dust growth.
The gas distribution on the phase diagram broadly follows the classical ``S-shape'' curve 
as determined by the thermal balance between radiative cooling and heating.
Hot gas with $T>10^5$~K is generated by SN feedback
while the narrow line at $T = 10^4$~K is the signature of photoionization from massive stars.

Without dust growth,
$Z^\prime_{\rm d}$ in hot gas decreases due to sputtering.
However,
the highly-sputtered gas (shown in red) is concentrated in the 
relatively high-density gas in the hot phase ($n = 0.1 - 1~{\rm cm}^{-3}$),
while the more diffuse hot gas is only weakly sputtered.
This results from the density dependence in the sputtering rate (see Eq.~\ref{eq:t_sput}).
There is a line of red points at the lowest temperatures ($\lesssim 30$~K)
whose origin is still unclear.
Nonthermal sputtering only operates when the dust-gas relative velocity is above 30 ${\rm km~s^{-1}}$,
which is unlikely for the cold gas.
We interpret this as gas particles whose associated dust was destroyed in SN shocks at an earlier time 
cool down to lower temperatures due to reduced photoelectric heating.

The situation becomes quite different if dust growth is included.
Firstly,
as already shown in Fig.~\ref{fig:nH_xi},
$Z^\prime_{\rm d}$ in high-density gas ($n>10^3~{\rm cm}^{-3}$) is strongly enhanced 
as this is the place where dust growth occurs.
Furthermore,
$Z^\prime_{\rm d}$ in the hot gas is slightly enhanced except for the densest and hottest region
where sputtering is most efficient.
This indicates that the high-$Z^\prime_{\rm d}$ dense gas where star formation occurs 
is dispersed by the subsequent stellar feedback and mixes with the ISM.
As sputtering only slightly decreases $Z^\prime_{\rm d}$,
the net effect is that the hot gas is ``dust-enriched''.
This is qualitatively similar to metal enrichment in supernova remnants,
where the SN ejecta of high metallicity mix with the ISM and increase its metallicity.

Observationally,
it is more straightforward to measure the projected DGR
$Z_{\rm d,proj} = \Sigma_{\rm d} / \Sigma_{\rm g}$
rather than the local $Z^\prime_{\rm d}$.
Fig.~\ref{fig:Sigma_vs_DGR} shows the normalized projected DGR 
$Z^\prime_{\rm d,proj}\equiv Z_{\rm d,proj} / Z_{\rm d,MW}$ 
as a function of the gas surface density ($\Sigma_{\rm g}$) 
with a pixel size $l_{\rm p} = 3$~pc
for our fiducial model (blue solid line)
and the model without dust growth (orange dashed line).

Without dust growth,
the projected DGR is fairly homogeneous everywhere.
Even in the SN bubbles ($\Sigma_{\rm g} \lesssim 1~{\rm M_\odot~pc^{-2}}$)
where sputtering occurs,
$Z_{\rm d}^{\rm \prime proj}$ is only 20\% lower.
Therefore,
sputtering alone is insufficient to generate DGR variation.
If dust growth is included,
we see significant DGR variation which can be broadly divided into three regimes: 
(1) the compact gas clumps ($\Sigma_{\rm g} \gtrsim 100~{\rm M_\odot~pc^{-2}}$) 
where 
$Z_{\rm d}^{\rm \prime proj}$ increases sharply with $\Sigma_{\rm g}$
as a direct consequence of the density-dependent dust growth.
(2) the diffuse ISM ($\Sigma_{\rm g} \approx 1 - 100~{\rm M_\odot~pc^{-2}}$) 
where 
$Z_{\rm d}^{\rm \prime proj}$ increases slowly with $\Sigma_{\rm g}$,
reflecting the large-scale DGR variation (e.g., radial gradient) as
the high-DGR gas clumps mix with the diffuse ISM
and
(3) the SN bubbles ($\Sigma_{\rm g} \lesssim 1~{\rm M_\odot~pc^{-2}}$) where $Z_{\rm d}^{\rm \prime proj}$ is 
enhanced by roughly a factor of two.

\subsubsection{Dust in galactic outflows}

\begin{figure*}
	\centering
	\includegraphics[trim = 1.8cm 0cm 1.8cm 0cm, clip,width=0.99\linewidth]{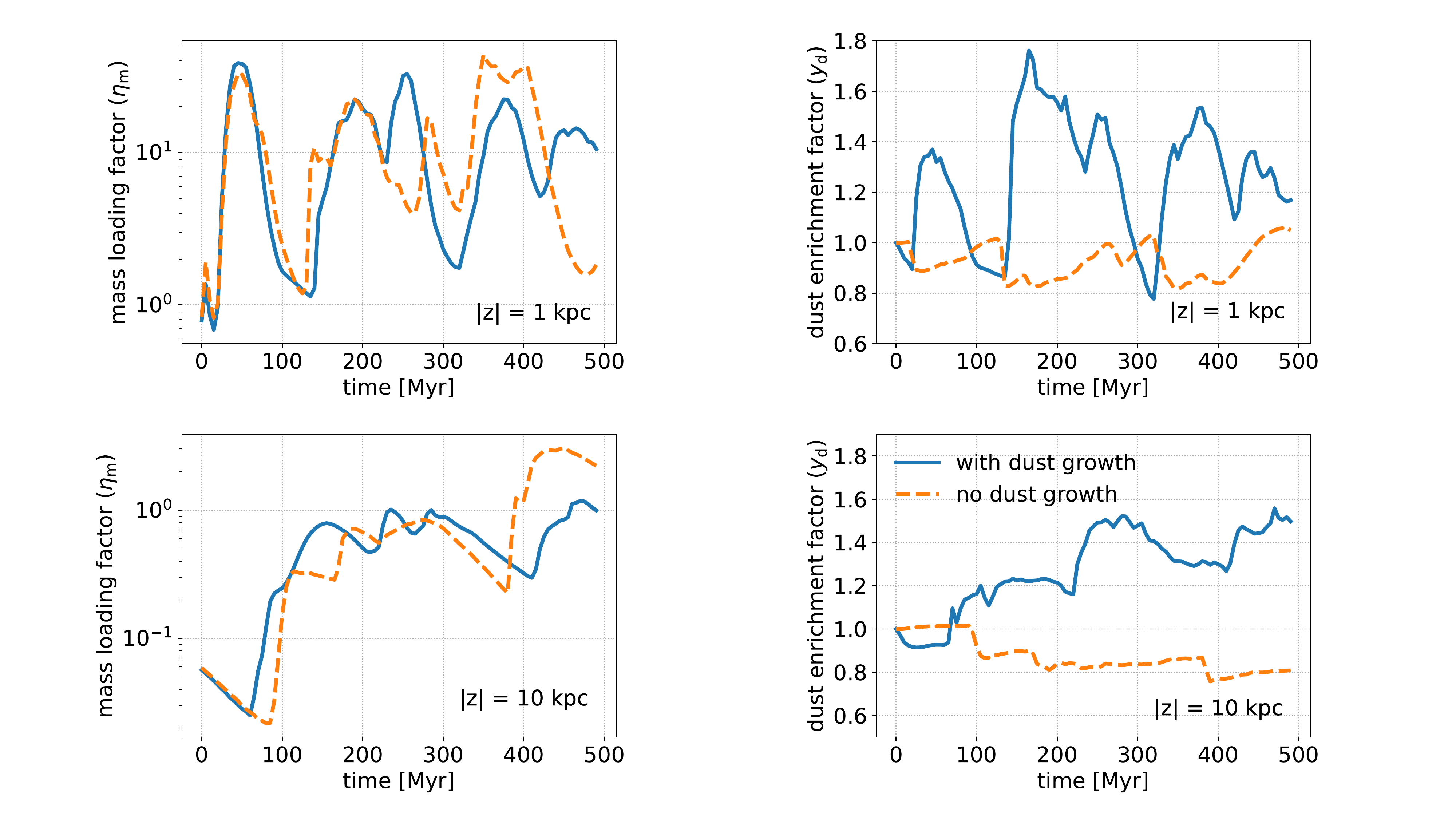}
	\caption{
		Time evolution of the
		mass loading factor (left panels)
		and the 
		dust enrichment factor (right panels)
		of the galactic outflows
		measured at $|z| = 1$~kpc (top panels) and $|z| = 10$~kpc (bottom panels).
		Dust growth leads to dust-enriched outflows.
	}
	\label{fig:OFR_dust}
\end{figure*}

Observations suggest that dust exists in the intergalactic medium far away from galaxies \citep{Menard2010, Peek2015}.
In this section,
we quantify the SN-driven outflow rates from our simulated galaxy.

We define the mass outflow rate for gas as 
\begin{equation}
	\dot{M}^{\rm out}_{\rm g} = \int_{S} \rho {\bf v} \cdot \hat{n} dA,
\end{equation}
where  $\rho$ is the gas density, ${\bf v}$ is the gas velocity,
$\hat{n}$ is the outward unit normal vector of the area $dA$ 
and $S$ is the surface where we measure the outflow rate.
Similarly,
the mass outflow rate for dust is defined as 
\begin{equation}
	\dot{M}^{\rm out}_{\rm d} = \int_{S} Z_{\rm d} \rho {\bf v} \cdot \hat{n} dA.
\end{equation}
In this work,
we measure the outflow rates 
at $|z| = z_{\rm out}$~kpc parallel to the mid-plane of the galactic disk.
We adopt two choices of $z_{\rm out} $: 1~kpc and 10~kpc.
Following \citet{Hu2019},
the discretized outflow rate for gas and dust can be expressed as
\begin{eqnarray}
	\dot{M}^{\rm out}_{\rm g}  &=& \sum_{ (z v_z)_{i}>0} \dfrac{ (m_{\rm g} v_{z} )_i}{dz}, \\
	\dot{M}^{\rm out}_{\rm d}  &=& \sum_{ (z v_z)_{i}>0} \dfrac{ (m_{\rm g} v_{z} Z_{\rm d} )_i }{dz}, 
\end{eqnarray}
where the subscript $i$ represents the particle index,
$v_{z}$ is gas velocity in the vertical direction,
and $dz = 0.1 z_{\rm out}$ is the thickness of the measuring plane.
The summation is over particles with $z v_z > 0$ (i.e., outflowing gas) within $z =  z_{\rm out} \pm 0.5 dz$
and $z =  -z_{\rm out} \pm 0.5 dz$.

We define the mass loading factor as  
\begin{eqnarray}
	\eta_{\rm m} (t) \equiv \frac{ \dot{M}_{\rm g}^{\rm out} (t) }{\rm \overline{SFR}}
\end{eqnarray}
where 
${\rm \overline{SFR}}$ is the time-averaged star formation rate over the entire simulation (500 Myr).
We take the averaged SFR instead of the instantaneous SFR for normalization
as there is a time delay between the star formation events 
and the associated outflowing gas arriving at the measuring planes.
Given the burstiness of the SFR,
if we took the instantaneous SFR for normalization,
$\eta_{\rm m}$ can be misleadingly high when the outflow rate is modest but the SFR is very low.

To quantify whether dust is preferentially expelled out of the galaxy,
we define the \textit{dust enrichment factor}
\begin{eqnarray}
	y_{\rm d} (t) \equiv \frac{ \dot{M}^{\rm out}_{\rm d} (t) }{ \dot{M}^{\rm out}_{\rm g} (t) Z_{\rm d}^{\rm ISM} (t) }
\end{eqnarray}
where $Z_{\rm d}^{\rm ISM}$ is the DGR in the ISM defined as $|z| < 0.5$~kpc.
Here we take the instantaneous DGR in the ISM, $Z_{\rm d}^{\rm ISM} (t)$, for normalization.
This is because, in contrast to the SFR,
the DGR in the ISM varies very slowly over time.
Note that the ratio $ \dot{M}^{\rm out}_{\rm d} /  \dot{M}^{\rm out}_{\rm g}$ 
is essentially the DGR of outflows weighted by the mass flux ($m_{\rm g} v_{z}$).

Fig.~\ref{fig:OFR_dust} shows 
the time evolution of 
$\eta_{\rm m}$ (left panels)
and 
$y_{\rm d}$ (right panels)
measured at $|z| = 1$~kpc (top panels) and $|z| = 10$~kpc (bottom panels).
The blue solid line shows our fiducial model while 
the orange dashed line shows the model without dust growth.

We first discuss the strength of outflows.
At $|z| = 1$~kpc,
$\eta_{\rm m}$ fluctuates strongly with time between 1 and 30
as a result of the bursty star formation and SN feedback.
At $|z| = 10$~kpc,
it drops by an order of magnitude to $\eta_{\rm m} = $~0.3 -- 1,
suggesting that a large fraction of outflows measured at $|z| = 1$~kpc 
is balanced by inflowing gas that falls back to the disk (i.e., fountain flows),
which is consistent with \citet{Hu2019}.
Dust growth has a negligible effect on $\eta_{\rm m}$ 
as it does not affect the thermal balance and the dynamics in the ISM significantly.
The difference between two models is likely due to the intrinsic stochasticity of star formation and feedback.

We now examine the dust content in outflows.
Without dust growth,
the dust enrichment factor is less than unity: $y_{\rm d} \sim 0.8$.
This is because
dust is sputtered in the shocked-heated gas in SNRs (see Fig.~\ref{fig:PD_dgr_growth})
that is then launched as outflows.
Sputtering only destroys $\sim 20\%$ of dust even when the outflows have traveled to $|z| = 10$~kpc.
The slightly lower $y_{\rm d}$ at $|z| = 1$~kpc 
does not mean that dust is created during its journey from $|z| = 1$~kpc to $|z| = 10$~kpc,
which is unlikely given the low gas densities.
Instead,
it is likely due to dilution by the entrained ISM which has $Z_{\rm d} = Z_{\rm d}^{\rm ISM}$ by definition 
and is expected to fall back as fountain flows rather than travel to $|z| = 10$~kpc.

The situation becomes very different once dust growth is included.
The outflows are now more dusty than the ISM,
with $y_{\rm d} \sim$~1.2 -- 1.5 at $|z| = 10$~kpc,
as the shocked-heated gas in SNRs is ``dust-enriched'' due to dust growth
(see Fig.~\ref{fig:PD_dgr_growth}).
Again,
this is analogous to metal enrichment from SN ejecta which leads to metal-enriched outflows.

\subsubsection{The effect of beam size}

\begin{figure*}
	\centering
	\includegraphics[width=0.99\linewidth]{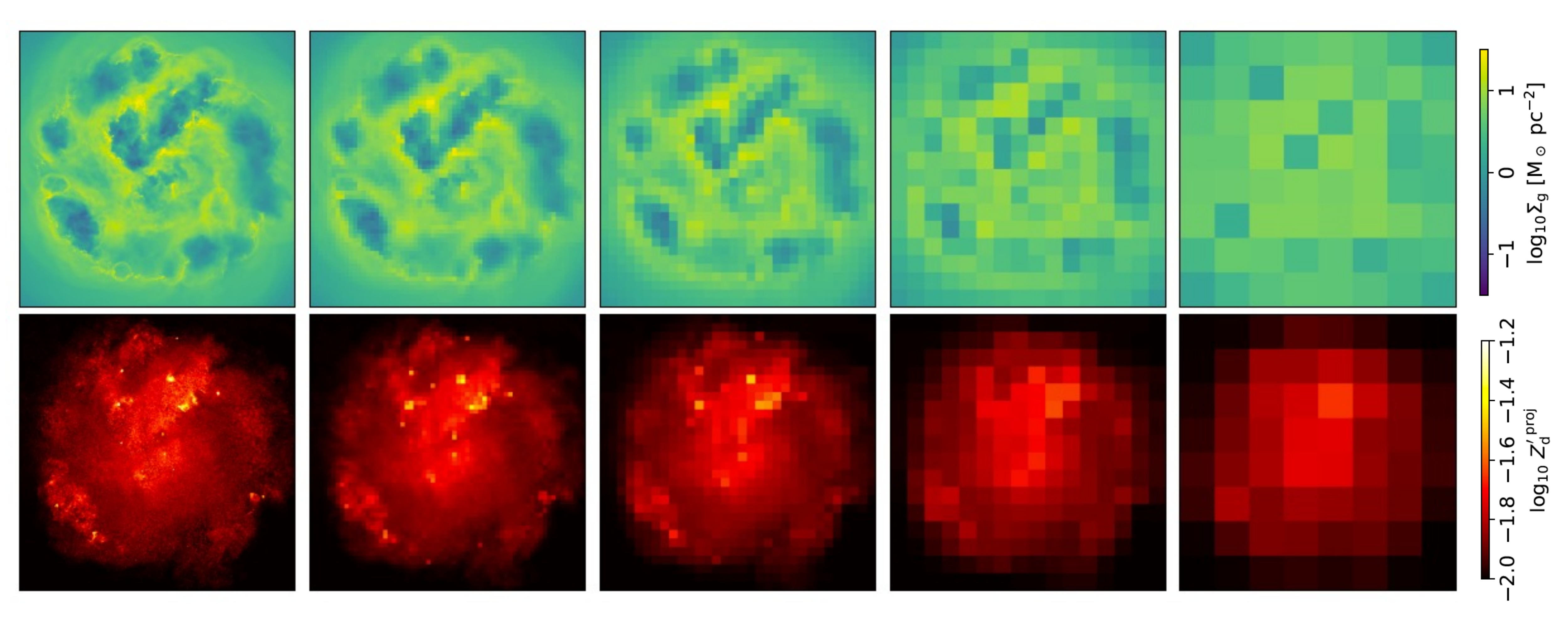}
	\caption{
		Images 
		of the gas surface density 
		(upper panels) and 
		the projected DGR
		(lower panels) at $t = 240$~Myr
		with $l_{\rm b} = $~24, 48, 96, 192, and 384~pc from left to right.
	}
	\label{fig:DGR_pix_15}
\end{figure*}

\begin{figure*}
	\centering
	\includegraphics[width=0.8\linewidth]{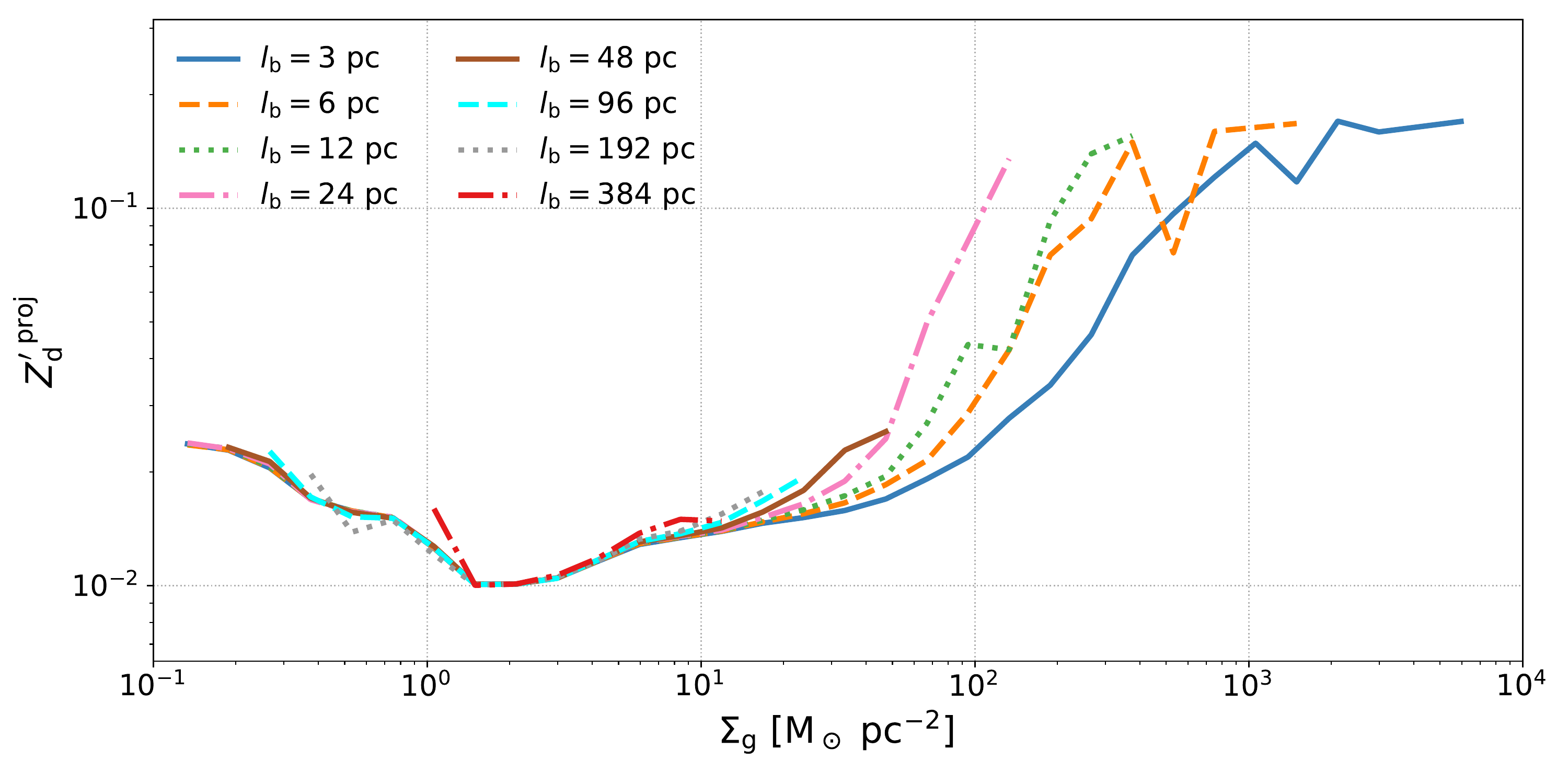}
	\caption{
		Same with Fig.~\ref{fig:Sigma_vs_DGR} but only for the fiducial model and
		with systematically coarser beam sizes ($l_{\rm b}$).
		The scatters are not shown for clarity.
		Beam averaging shifts the relationship to lower gas surface densities as 
		it smooths out the high-DGR dense gas.
	}
	\label{fig:pix_Sigma_vs_DGR}
\end{figure*}

Extragalactic observations often have a telescope beam size significantly coarser than 3~pc 
as adopted in Fig.~\ref{fig:Sigma_vs_DGR}.
To understand the effect of beam size,
we construct images of $\Sigma_{\rm g}$ and $\Sigma_{\rm d}$ 
for our fiducial model
at systematically coarser beam sizes of
$l_{\rm b} = $~3, 6, 12, 24, 48, 96, 192, and 384~pc.
For example,
\begin{align}
	\Sigma_{\rm g}\,(6~{\rm pc}) &= \frac{\int \Sigma_{\rm g}\,\mathrm{d}A}{\int \mathrm{d}A} = \frac{1}{2^2} \sum_{i=1}^{4} \Sigma_{{\rm g},i}\,(3~{\rm pc})~,\\
	\Sigma_{\rm d}\,(6~{\rm pc}) &= \frac{\int \Sigma_{\rm d}\,\mathrm{d}A}{\int \mathrm{d}A} = \frac{1}{2^2} \sum_{i=1}^{4} \Sigma_{{\rm d},i}\,(3~{\rm pc})~,
\end{align}
and, correspondingly, 
\begin{align}
	Z_{\rm d}^{\rm \prime proj}\,(6~{\rm pc}) = \frac{\Sigma_{\rm d}\,(6~{\rm pc}) }{ \Sigma_{\rm g}\,(6~{\rm pc}) Z_{\rm d,MW}}.
\end{align}

Fig.~\ref{fig:DGR_pix_15} shows the coarsened images
of $\Sigma_{\rm g}$ (upper panels) and $Z_{\rm d}^{\rm \prime proj}$ (lower panels) at $t = 240$~Myr
with $l_{\rm b} = $~24, 48, 96, 192, and 384~pc from left to right.
The compact dense gas with very high $Z^\prime_{\rm d}$ can be resolved reasonably well at $l_{\rm b} = $~24 pc.
As  $l_{\rm b}$ increases,
this high-$Z^\prime_{\rm d}$ gas is gradually smoothed out and the DGR is significantly diluted
by the diffuse ISM which has a much lower DGR.
At $l_{\rm b} = $~384 pc,
$Z_{\rm d}^{\rm \prime proj}$ becomes very uniform.
There is a slight radial gradient of $Z_{\rm d}^{\rm \prime proj}$
which reflects the large-scale radial distribution of the gas surface density,
similar to the metallicity gradient in galaxies caused by metal enrichment.

To be more quantitative,
Fig.~\ref{fig:pix_Sigma_vs_DGR} shows the relationship between 
$\Sigma_{\rm g}$  and $Z_{\rm d}^{\rm \prime proj}$ at various beam sizes
for all snapshots between $t = $~100 -- 500 Myr.
As $l_{\rm b}$ increases,
beam averaging smooths out the dense gas
such that both $\Sigma_{\rm g}$ and $\Sigma_{\rm d}$ decrease.
However,
$\Sigma_{\rm d}$ decreases less significantly
because $Z_{\rm d}^{\rm \prime proj}$ is significantly higher at high $\Sigma_{\rm g}$,
shifting the relationship leftward.
As a result,
the gas surface density above which $Z_{\rm d}^{\rm \prime proj}$ rises sharply decreases at larger $l_{\rm b}$. 
The observational implication is that
measurements of the $\Sigma_{\rm g}$ -- $Z_{\rm d}^{\rm \prime proj}$ relationship
must be compared at a similar beam size.

At even larger beam sizes ($l_{\rm b} \geq 92$~pc),
the sharply rising part disappears completely,
and the $\Sigma_{\rm g}$ -- $Z_{\rm d}^{\rm \prime proj}$ relationship becomes insensitive to $l_{\rm b}$.
This happens when the high-$Z_{\rm d}^\prime$ dense gas is completely diluted away by the diffuse ISM at large enough $l_{\rm b}$.
The slowly-rising part corresponds to the large-scale radial gradient of $Z_{\rm d}^{\rm \prime proj}$ as can be seen in Fig.~\ref{fig:DGR_pix_15}.

The $\Sigma_{\rm g}$ -- $Z_{\rm d}^{\rm \prime proj}$ relation likely depends
on the metallicity and the large-scale gas surface density.
We plan to conduct simulations of SMC- and LMC-like galaxies in the future
for direct comparison with high-resolution FIR and UV observations such as \citet{RomanDuval2017, RomanDuval2022}.

\newpage

\section{Discussion}\label{sec:discussion}

\subsection{Implications for the observationally derived DGR}\label{sec:obsDGR}

Observationally,
the galaxy-integrated DGR is often estimated using
\begin{equation}\label{eq:obsDGR}
	Z_{\rm d} = \frac{ M_{\rm d}}{M_{\rm H_2} + M_{\rm {\hi}}} 
	= \frac{ M_{\rm d}}{L_{\rm CO} \alpha_{\rm CO} + M_{\rm {\hi}}},
\end{equation}
where $M_{\rm {\hi}}$ is the total {\hi} mass from the 21-cm line 
and $M_{\rm d}$ is the total dust mas from the FIR continuum.
A major uncertainty is in the adopted $\alpha_{\rm CO}$.
As a result,
\citet{RemyRuyer2014} reported two versions of the metallicity--DGR relation
based on two different choices of $\alpha_{\rm CO}$,
one is a constant Milky Way value $\alpha_{\rm CO,MW}$
and the other is metallicity-dependent that scales as $Z^{\prime -2}$.
While the 
latter one is more frequently adopted in the literature,
our results suggest that 
the one based on the Milky Way conversion factor 
is actually more appropriate at low metallicity.
In fact,
the metallicity-dependent $\alpha_{\rm CO}$ strongly overestimates $M_{\rm H_2}$ 
which in turn underestimates $Z_{\rm d}$.

That said,
our adopted initial DGR $Z^\prime_{\rm d,0} = 0.01$ is motivated by the metallicity--DGR relation in \citet{RemyRuyer2014}
that assumes $\alpha_{\rm CO}\propto Z^{\prime -2}$,
which, as we just argued above, could be an underestimate.
We rerun our fiducial model in Appendix~\ref{app} with $Z^\prime_{\rm d,0} = 0.03$
and find $\alpha_{\rm CO} \sim 3 \alpha_{\rm CO,MW}$.

Furthermore,
our simulations showed,
consistent with previous studies in \citet{Hu2016,Hu2017},
that the molecular mass fraction is extremely small ($F_{\rm H_2}\sim 10^{-4}$) in dwarf galaxies with $Z^\prime=0.1$
such that $M_{\rm H_2}$ contributes negligibly to the total gas mass.
This is a natural consequence of the long H$_2$ formation time $t_{\rm H_2}\sim 1~{\rm Gyr}(n Z^\prime_{\rm d})^{-1}$.
With our adopted $Z^\prime_{\rm d} = 0.01$,
gas at $n = 100~{\rm cm}^{-2}$ takes $t_{\rm H_2} \sim 1~{\rm Gyr}$ to form H$_2$,
which is orders of magnitude longer than the Myr-scale dynamical time in the ISM.
Therefore,
the H$_2$ abundance is primarily limited by the dynamical time \citep{Glover2011, Hu2021a}.
The situation
remains qualitatively similar
even if we adopted $Z^\prime_{\rm d} = 0.1$
(i.e., a linear metallicity--DGR relation) as shown in \citet{Hu2016} who found $F_{\rm H_2}\sim 10^{-3}$.
Consequently,
the DGR should be observationally estimated simply by $Z_{\rm d} = M_{\rm d} / M_{\rm {\hi}}$ at low metallicity,
and the uncertainty in $\alpha_{\rm CO}$ is irrelevant.

\subsection{Implications for the intergalactic dust}\label{sec:IGMdust}

Observations of reddening
suggest that dust exists in the intergalactic medium 
20 kpc to several Mpc away from galaxies \citep{Menard2010, Peek2015},
whose origin is still poorly understood.
One of the possible scenarios is dust entrained in galactic outflows \citep{Aguirre1999, Bianchi2005}.
\citet{Kannan2021} 
conducted simulations of isolated galaxies with similar properties of the Milky Way and LMC
coupled with a dust evolution model.
They found that outflows are able to entrain dust in the fountain flows
that circulate around the galaxies within a few kpc.
However,
they found that dust cannot be expelled to the outer part of the halos,
which is in contrast to our case where outflows at 10 kpc are still dust-enriched (rather than depleted)
and are expected to eventually escape the halo \citep{Hu2019}.
This might indicate that dwarf galaxies are preferable sites to pollute
the intergalactic medium with dust as they have low gravitational potential wells and
the lack of hot gaseous halos around them prevents further destruction via thermal sputtering.
On the other hand,
the difference could also arise from numerics as 
the resolution in \citet{Kannan2021} is $10^3~{\rm M_\odot}$ which was presumably 
too coarse to resolve the dense gas where dust growth is most efficient.
Resolved simulations of LMC-like galaxies have recently been conducted by \citet{Steinwandel2022},
and a systematic study of outflows across a range of galaxy masses will be valuable to shed light on this topic.


\subsection{Neglected physics}

Our dust evolution model does not include dust production in SN ejecta,
which has been observed \citep{Wooden1993, Indebetouw2014}
and 
might be major source of dust production in the early Universe 
when there was not enough time for AGB stars to kick in.
It is still an open question 
how much dust will eventually survive once the reverse shock hits back
which depends sensitively on local gas properties 
\citep{Bianchi2007, Micelotta2016, Kirchschlager2019, Priestley2022}.
Including dust production in SNe is unlikely to affect the DGR in dense gas and the ISM chemistry, 
but it could make the shock-heated hot gas and galactic outflows even dustier.

In addition, our dust model assumes a fixed ``MRN'' grain-size distribution 
and neglects processes that can vary the grain size
such as shattering and coagulation.
As the timescales for sputtering and dust growth both depend linearly on grain size,
the dust production/destruction rate would be affected
if the actual grain-size distribution deviates significantly from MRN.
In addition,
the grain-size distribution may also affect the H$_2$ formation rate on dust
as well as dust shielding \citep{Jonkheid2006, Romano2022}.
Including the evolution of grain-size distribution 
is therefore a highly desirable extension for future work.







\section{Summary}\label{sec:summary}

We have presented the first ISM-resolved galaxy scale simulations
coupled with time-dependent hydrogen chemistry and dust evolution.
Our aim is to understand the ISM chemistry and dust properties at low metallicity,
a condition expected to be common for galaxies in the very early Universe observed by JWST.
Our simulated galaxy is similar to the WLM dwarf galaxy,
which has a metallicity of 0.1 $Z_\odot$ and has detected {\coone} emission.
We adopt a mass resolution of $1~{\rm M_\odot}$ per gas particle
which corresponds to a spatial resolution $\sim 0.2$~pc
to properly resolve the compact CO cores.
We post-process the simulation snapshots with a detailed chemistry network 
to accurately model the C$^+$/{\ci}/CO transitions, 
taking the time-dependent abundances of H$_2$ and H$^+$ from the simulations as input parameters.
Our main findings can be summarized as follows.

\begin{enumerate}

	\item 
	Our fiducial simulation successfully reproduces both the 
	observed SFR and {\coone} luminosity ($L_{\rm CO}$) in the WLM dwarf galaxy (Fig.~\ref{fig:time_H2COSFR}).
	$L_{\rm CO}$ can only be
	reproduced if dust growth in the ISM is included,
	as otherwise dust shielding would be insufficient to protect CO 
	from being photodissociated by FUV radiation, 
	suppressing $L_{\rm CO}$ by more than two orders of magnitude (Fig.~\ref{fig:nH_xi}).

	\item 
	The predicted total H$_2$ fraction is extremely low ($\sim 10^{-4}$) either with or without dust evolution
	due to the long H$_2$ formation time.
	This leads to very little CO-dark H$_2$ gas and 
	a CO-to-H$_2$ conversion factor (excluding helium) 
	$\alpha_{\rm CO} = 4.63~{\rm M_\odot~pc^{-2}~(K~km~s^{-1})^{-1}}$
	which is only slgithly higher than the Milky Way value despite the low metallicity (Table~\ref{tab:sumstats}).
	Observationally inferred dust-to-gas ratio (DGR) is underestimated at low metallicity
	if assuming a very steep metallicity-dependent $\alpha_{\rm CO}$ (Section~\ref{sec:obsDGR} and Appendix~\ref{app}).

	\item 
	Dust growth is the primary driver of the spatial variation of DGR in the ISM.
	Dust growth significantly increases the DGR in cold, star-forming clouds.
	Subsequent SN feedback disperses the high-DGR gas without significantly destroying the dust,
	leading to elevated DGR in the diffuse gas associated with supernova remnants,
	qualitatively similar to metal enrichment (Figs.~\ref{fig:PD_dgr_growth} and \ref{fig:Sigma_vs_DGR}).
	As a result,
	galactic outflows are about 20 -- 50\% dustier than the ISM (Fig.~\ref{fig:OFR_dust}).
	Dwarf galaxies therefore could be a source of the intergalactic dust (Section~\ref{sec:IGMdust}).

	\item 
	The projected DGR increases with gas surface density due to dust growth.
	The relationship between these two quantities varies with the telescope beam size 
	as coarse beams smooth out the high-DGR gas clumps
	(Figs.~\ref{fig:DGR_pix_15} and \ref{fig:pix_Sigma_vs_DGR}).
	Observational measurements should therefore be compared at a similar beam size.


\end{enumerate}

\section*{Acknowledgments}
We thank the anonymous referee for their constructive comments that improved our manuscript.
C.Y.H. acknowledges support from the DFG via German-Israel Project Cooperation grant STE1869/2-1 GE625/17-1
and NASA ATP grant 80NSSC22K0716.
A.S. thanks the Center for Computational Astrophysics (CCA) of the Flatiron Institute,
and the Mathematics and Physical Science (MPS) division of the Simons Foundation for support.
All simulations were run on the Raven and Cobra supercomputers at the Max Planck Computing and Data Facility (MPCDF).

\bibliography{literatur}{}
\bibliographystyle{aasjournal}

\appendix
\section{Effect of adopted initial DGR}\label{app}
In this section,
we rerun our fiducial model with an initial DGR $Z^\prime_{\rm d} = 0.03$ 
(instead of $Z^\prime_{\rm d} = 0.01$ as in our fiducial model) for 160 Myr.

\begin{figure*}
	\centering
	\includegraphics[trim = 1.8cm 0cm 1.8cm 0cm, clip, width=0.85\linewidth]{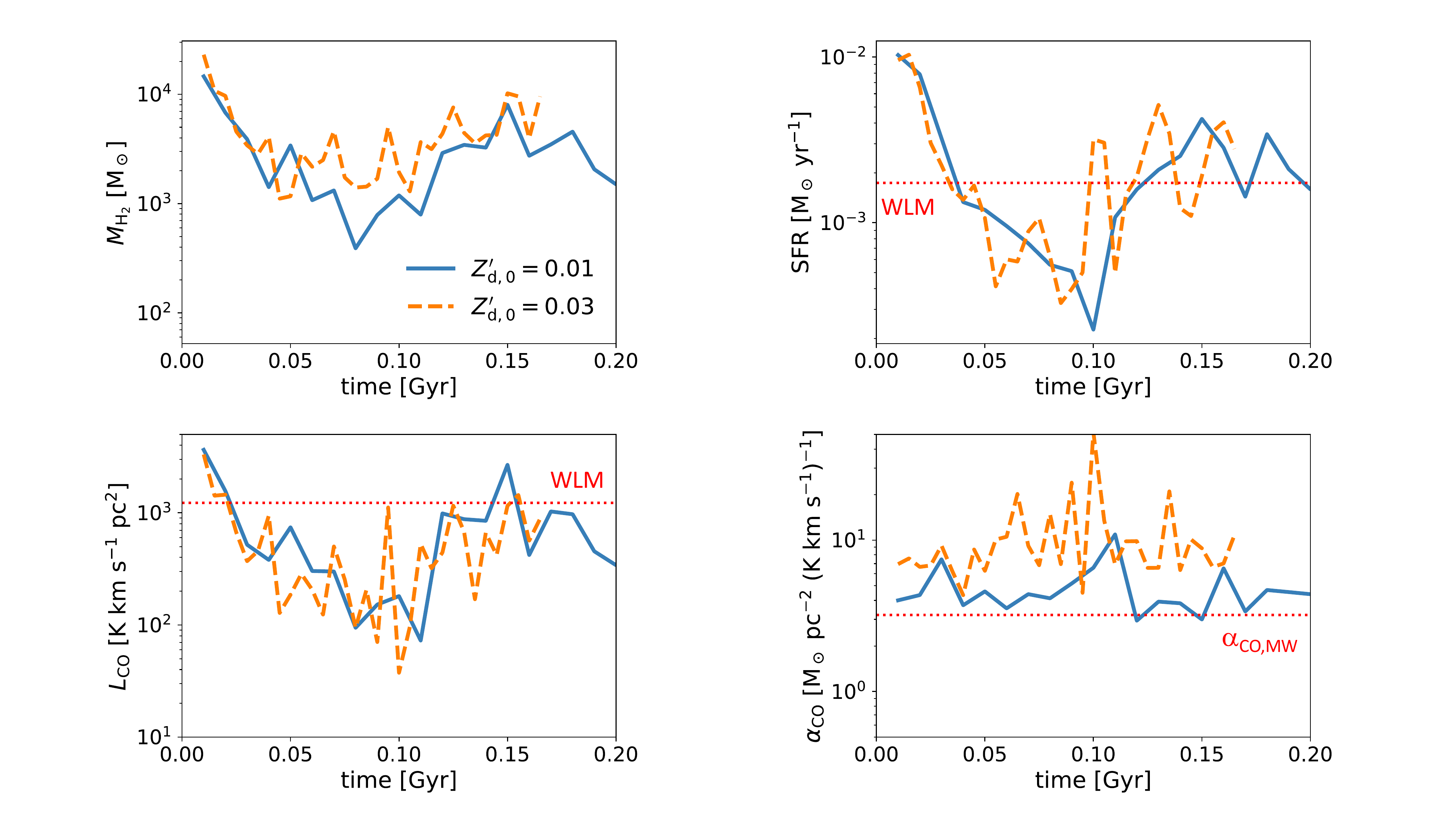}
	\caption{
		Same with Fig.~\ref{fig:time_H2COSFR} but comparing models with different initial DGR: $Z^\prime_{\rm d} = 0.01$ (blue solid) vs. $Z^\prime_{\rm d} = 0.03$ (orange dashed).
	}
	\label{fig:time_H2COSFR_D0p03}
\end{figure*}

Fig.~\ref{fig:time_H2COSFR_D0p03}
shows the time evolution of $M_{\rm H_2}$, SFR, $L_{\rm CO}$, and $\alpha_{\rm CO}$ similar to Fig.~\ref{fig:time_H2COSFR}
for the fiducial model with $Z^\prime_{\rm d} = 0.01$ (blue solid) and $Z^\prime_{\rm d} = 0.03$ (orange dashed), respectively. 
The SFR and $L_{\rm CO}$ are very similar in both models,
but $M_{\rm H_2}$ is slightly higher in the case of $Z^\prime_{\rm d} = 0.03$, 
leading to $\alpha_{\rm CO}\sim 3 \alpha_{\rm CO,MW}$.
We note that $M_{\rm H_2}$ still contributes a negligible fraction of the total gas mass ($M_{\rm H_2} \ll M_{\hi}$)
and thus it can be ignored when deriving $Z_{\rm d}$ observationally with Eq.~\ref{eq:obsDGR}.

\begin{figure*}
	\centering
	\includegraphics[width=0.75\linewidth]{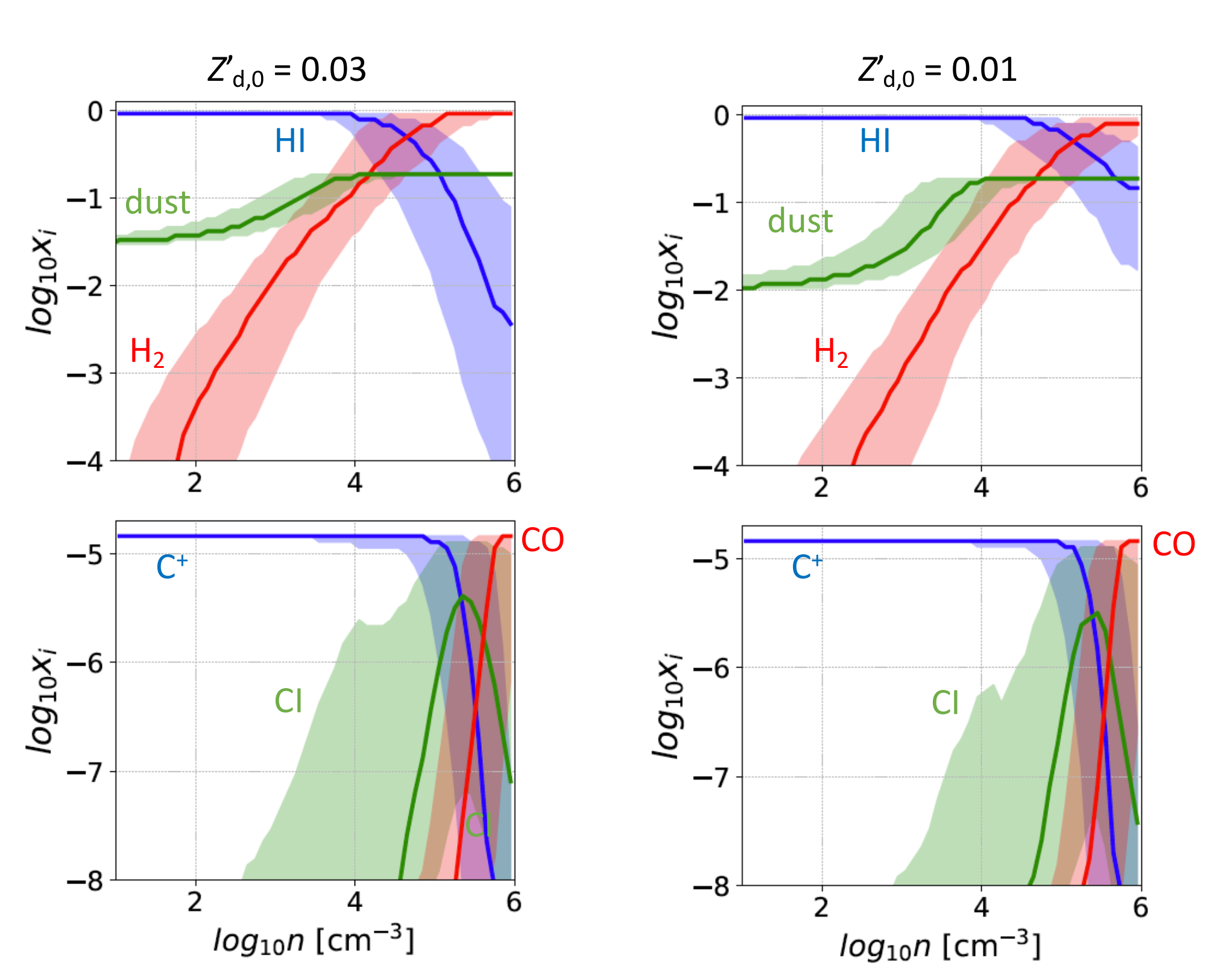}
	\caption{
		Same with Fig.~\ref{fig:nH_xi}, but comparing models with different initial DGR: $Z^\prime_{\rm d} = 0.03$. vs. $Z^\prime_{\rm d} = 0.01$.
	}
	\label{fig:nH_xi_D0p03}
\end{figure*}

Fig.~\ref{fig:nH_xi_D0p03} shows $Z^\prime_{\rm d}$ and the chemical abundances of {\hi}, H$_2$, C$^+$, C, and CO in a similar way as in Fig.~\ref{fig:nH_xi}.
H$_2$ is more abundant in the case of $Z^\prime_{\rm d} = 0.03$ due to more efficient H$_2$ formation and H$_2$ self-shielding.
On the other hand,
the transition of C/CO is relatively insensitive to $Z^\prime_{\rm d} = 0.03$.
This can be interpreted as dust shielding is mostly originated from the dense gas in the vicinity of the CO cores
where the DGR is saturated in both cases.
As such, the three times difference in $Z^\prime_{\rm d}$ in the diffuse ISM has a negligible effect on the C/CO transition,
which explains the similar $L_{\rm CO}$ in both models.

\end{document}